\documentclass[reprint,
nofootinbib,
 amsmath,amssymb,
 aps,
prstab
]{revtex4-1}

\usepackage{graphicx} 
\usepackage{subcaption}
\usepackage{dcolumn}
\usepackage{bm}
\usepackage{xcolor}
\usepackage{natbib} 
\usepackage[nottoc,numbib]{tocbibind}
\usepackage{amsmath}
\usepackage{caption}
\usepackage{soul}
\usepackage[normalem]{ulem}
\usepackage{ulem}
\usepackage{changes} 
\usepackage{siunitx}
\DeclareUnicodeCharacter{2060}{}
\DeclareUnicodeCharacter{00A0}{~}

\usepackage{lipsum}

\usepackage[%
  colorlinks=true,
  urlcolor=blue,
  linkcolor=blue,
  citecolor=blue
]{hyperref}
\usepackage{stackengine}
\usepackage{overpic}
\usepackage{pict2e}
\usepackage{tikz}
\usepackage{comment}

\usepackage{amsmath}

\usepackage{tabularx}
\usepackage{xcolor}
\usepackage{textcomp} 

\makeatletter
\newcommand\thefontsize[1]{{#1 The current font size is: \f@size pt\par}}
\newcommand\thefontsizeHere{{The current font size is: \f@size pt\par}}
\makeatother

\usepackage{tikz,xcolor,hyperref}

\definecolor{lime}{HTML}{A6CE39}
\DeclareRobustCommand{\orcidicon}{%
	\begin{tikzpicture}
	\draw[lime, fill=lime] (0,0) 
	circle [radius=0.16] 
	node[white] {{\fontfamily{qag}\selectfont \tiny ID}};
	\draw[white, fill=white] (-0.0625,0.095) 
	circle [radius=0.007];
	\end{tikzpicture}%
}

\foreach \x in {A, ..., Z}{%
	\expandafter\xdef\csname orcid\x\endcsname{\noexpand\href{https://orcid.org/\csname orcidauthor\x\endcsname}{\noexpand\orcidicon}}
}

\begin{document}

\preprint{APS/123-QED}

\title{RF cavity design and beam loading effects in the rectilinear cooling channel for a muon collider}

\newcommand{\orcidauthorA}{0000-0001-7594-5840} 

\newcommand{\orcidauthorE}{0000-0001-9759-2572} 

\newcommand{\orcidauthorB}{0009-0006-5828-5098} 

\newcommand{\orcidauthorC}{0000-0001-6336-3528} 

\newcommand{\orcidauthorD}{0000-0001-7358-9686} 

\author{C.~Barbagallo\orcidA{}}
\thanks{Corresponding author: carmelo.barbagallo@cern.ch}
\affiliation{European Organization for Nuclear Research (CERN), Geneva, Switzerland}

\author{A.~Grudiev\orcidE{}}
\affiliation{European Organization for Nuclear Research (CERN), Geneva, Switzerland}

\author{D.~Merenich\orcidB{}}
\affiliation{Northern Illinois University, DeKalb, Illinois 60115, USA}

\author{X.~Lu\orcidC{}}
\thanks{Also at Argonne National Laboratory, Lemont, Illinois 60439, USA}
\affiliation{Northern Illinois University, DeKalb, Illinois 60115, USA}

\author{T.~Luo\orcidD{}}
\affiliation{Lawrence Berkeley National Laboratory, Berkeley, California 94720, USA}


\begin{abstract}

High-gradient normal-conducting RF cavities are crucial components of the rectilinear cooling channel in a muon collider, providing the accelerating field necessary to sustain the rapid six-dimensional ionization cooling essential to achieve high luminosity. To reach this goal, a muon collider requires approximately $10^{12}$ muons per bunch. At such high beam intensities, beam loading induced by bunch-driven wakefields can significantly reduce the cavity accelerating gradient and compromise beam stability. This paper focuses on the conceptual RF design and beam-induced effect analysis of copper cavities with so-called beam windows developed to meet the beam dynamics requirements of the rectilinear muon cooling channel. We first present the design and optimization of the cavities, evaluating their RF performance and power requirements. Thereafter, we analyze cavity wakefields and beam loading effects, proposing a new numerical method for their compensation, which takes the transverse distribution of the beam into account.  

\end{abstract}

\maketitle

\section{Introduction}

The muon collider design represents a novel and compact high-energy lepton accelerator concept to produce, cool, accelerate, and collide counter-propagating single-bunch muon beams of opposite charge~\cite{Long_2021, accettura2025muon}. A high-energy muon collider would combine the unparalleled energy reach characteristic of proton colliders with the precision typical of electron-positron rings, enabling direct exploration of the Standard Model and offering substantial discovery potential beyond it. The reference design for the muon collider targets 10~TeV center-of-mass collisions and an integrated luminosity of 10 $\mathrm{ab}^{-1}$~\cite{Accettura_2023}. This ambitious goal, with a first-stage facility envisioned to begin operation around 2050, has motivated the scientific community to expand the boundaries of accelerator R\&D over the past few decades. The conceptual design of the muon collider was pioneered in earlier feasibility studies, such as those presented in the CERN 99-02 report in 1999~\cite{CERN9902}. These concepts were later advanced by the U.S. Muon Accelerator Program (MAP) up to 2017~\cite{MAP_Ref}. The global effort is now coordinated by the International Muon Collider Collaboration (IMCC)~\cite{accettura2024imcc}.

A primary technical challenge for a muon collider is the large emittance of the muon beam produced at the source, which renders it unsuitable for high-luminosity collisions unless the emittance is reduced by several orders of magnitude through a cooling process~\cite{PhysRevAccelBeams.28.041003}. Furthermore, owing to the short muon lifetime (2.2~\textmu s in the rest frame), the cooling process must occur rapidly before muon decay significantly reduces the beam intensity. Six-dimensional~(6D) ionization cooling~\cite{neuffer2004introduction} is currently considered the only feasible method for reducing the emittance of muon beams in the present muon collider baseline. 

\begin{figure}[!htb]
   \centering
   \includegraphics*[width=1\columnwidth]{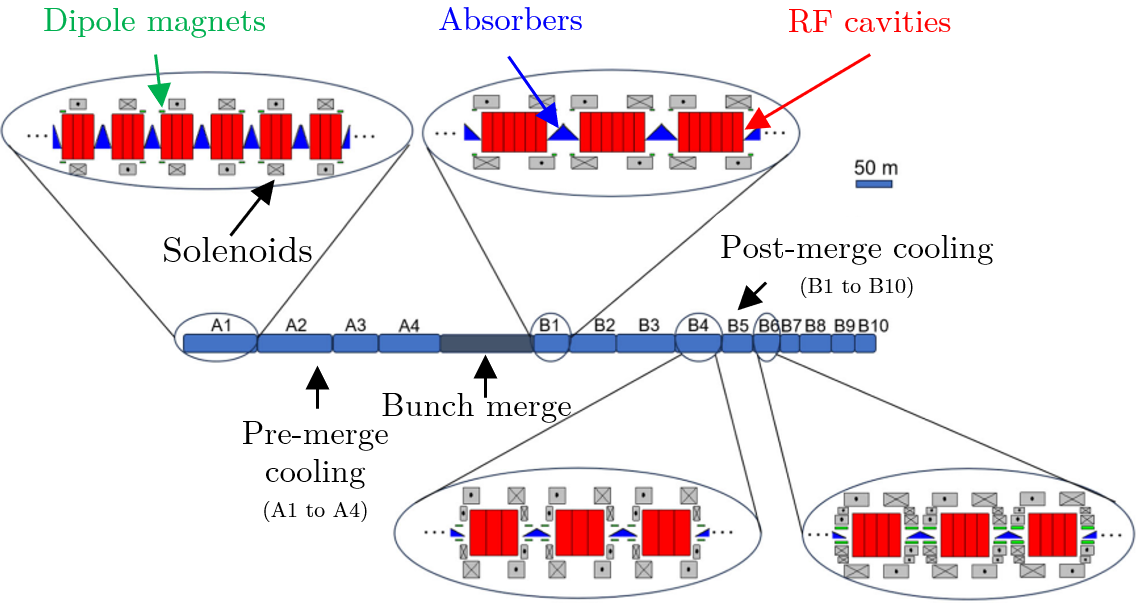}
   \vspace{0.1\baselineskip}
   \caption{Schematic layout of the rectilinear muon cooling channel. The channel is divided into a pre-merge section (stages A1 to A4), a bunch merging system, and a post-merge section (stages B1 to B10). Zoomed views illustrate the periodic lattice cell configurations, depicting the integration of RF cavities (red), absorbers (blue triangles), dipole magnets (green blocks), and focusing solenoids (gray blocks). Adapted from~\cite{PhysRevAccelBeams.28.041003}.}
   \label{fig:Rectilinear_cooling_channel_layout}
\end{figure}

The current baseline cooling lattice design is a rectilinear channel based on the MAP concept~\cite{Stratakis_2015}, which has recently been reviewed and optimized by the IMCC~\cite{PhysRevAccelBeams.28.041003}. A schematic layout of the rectilinear cooling channel is illustrated in Fig.~\ref{fig:Rectilinear_cooling_channel_layout}

The updated lattice design consists of two cooling channels: one before (Rectilinear A) and one after (Rectilinear B) the bunch merging system. In the following, Rectilinear A and B are referred to as the pre-merging and post-merging sections, respectively. Each section is divided into multiple stages, four for Rectilinear A and ten for Rectilinear B, with each stage consisting of a series of identical periodic units, denoted as cooling cells. Each cooling cell includes liquid hydrogen (LH$_2$) wedge absorbers to provide 4D transverse cooling by reducing total beam momentum via ionization energy loss, and radio-frequency (RF) pillbox-type cavities to restore the longitudinal momentum lost in the absorbers. Additionally, solenoids of opposite polarity are used to achieve tight focusing at the absorbers to minimize emittance growth from multiple Coulomb scattering, and dipole magnets are employed to generate dispersion. This dispersion, combined with the wedge shape of the absorbers, enables emittance exchange between the transverse and longitudinal planes, thereby accomplishing full 6D cooling. After undergoing sufficient cooling in Rectilinear A, an initial train of 21 bunches is merged into a single bunch by the bunch merging system. The merged bunch then enters the post-merging section (Rectilinear B), where further cooling is applied to reduce both its transverse and longitudinal emittance. 

To achieve the target beam parameters for the muon collider, the rectilinear cooling channel must reduce the 6D emittance by several orders of magnitude down to the sub-millimeter normalized emittances defined in the baseline design~\cite{PhysRevAccelBeams.28.041003}. To compensate for the ionization energy loss across all cooling stages, the cooling channel requires a cumulative accelerating voltage on the order of several gigavolts, provided by thousands of high-gradient RF cavities. The RF system envisioned for the rectilinear cooling channel requires $3{,}662$ pillbox-type RF cavity cells, operating at either 352~MHz or 704~MHz, depending on the cooling stage. Utilizing separate, parallel cooling lines for $\mu^+$ and $\mu^-$ muon beams, the total number of cavities in the proposed cooling channel is $7{,}324$.
These cavity cells are independently phased and accelerate muons in the TM$_{010}$ mode in a multi-Tesla magnetic field~\cite{Bowring_2020}. Additionally, they are sealed at their extremities with low-atomic-number metallic foils to enhance the achievable accelerating gradient in such a high magnetic field. These foils also lower the surface electric field to mitigate the risk of RF breakdown~\cite{wangler2008rf}, i.e., the sudden absorption of RF power initiated by field-emitted electrons at the cavity walls, while ensuring adequate beam transmission through the different stages. Significant efforts have been dedicated to the design of such RF cavities~\cite{li2001design}, culminating in the construction and testing of an 805 MHz cavity with beryllium (Be) windows at Fermilab~\cite{Li:EPAC02-2160}.

The intense muon bunches required for the collider, which reach an intensity on the order of $10^{12}$ muons per bunch by the end of the rectilinear cooling channel, can induce significant wakefields within the cavities~\cite{taylor2025mucol}. These beam-induced fields can reduce the effective cavity accelerating gradient and degrade beam quality. Simulations of short-range wakefields in a flat Be-windowed pillbox cavity generated by a sub-relativistic beam bunch were previously performed by H. Wang et al.~\cite{wang2001short} in the context of the muon ionization cooling channel design for a neutrino factory in the USA~\cite{alsharo2003recent}.
Subsequently, the effects of beam loading and wakefields in 325 and 650~MHz pillbox cavities for muon ionization cooling were investigated by M. Chung et al.~\cite{chung2014effects} for the MAP study~\cite{delahaye2013enabling}. However, these studies did not specifically focus on a detailed separation of the space-charge and resonant wakefield effects, nor their corresponding beam-induced voltage contributions within the total beam loading. In addition, strategies for beam-loading compensation were beyond their scope.

In this study, guided by the beam dynamics specification from the most recent muon cooling channel lattice~\cite{PhysRevAccelBeams.28.041003} and drawing inspiration from previous designs of RF cavities for the Muon Ionisation Cooling Experiment (MICE) project~\cite{li2001design, Li:EPAC02-2160, li2005progress, rimmer2025fabrication, li2006201}, we designed and optimized elliptical copper (Cu) RF cavities with irises closed by Be windows for each stage of the rectilinear muon cooling channel and studied the beam loading effects. In Sec.~\ref{sec:2}, we discuss the design and optimization of the cavities, aimed at preserving the on-axis electric fields to enhance muon acceleration, reduce the power consumption, and minimize surface electric fields to mitigate RF breakdown. We present RF performance metrics evaluated for both pre- and post-merging stages, together with a discussion of the advantages and limitations of closing the irises with Be windows. In Sec.~\ref{sec:3} and Sec.~\ref{sec:4}, we analyze wakefields and beam loading effects, respectively. We propose numerical methods that address the separation of space-charge contributions from purely cavity-dependent effects, and present strategies for their compensation. The dependence of the longitudinal wakefields on the transverse beam offset is analyzed in detail for cavities with and without Be windows. In addition, we present a new analytical method to include the finite transverse size of the beam in the wakefield calculations. This work provides further insights into Be-windowed cavities, where confined fields behave differently from those in conventional RF cavities, and the transverse dependence of the beam-cavity interaction is significant. We then present a benchmark comparing the resonant wake potential reconstructed from eigenmode solutions with the post-processed wake obtained from transient Particle-In-Cell (PIC) simulations. All electromagnetic simulations were conducted for the designed cavities in CST Studio Suite\textsuperscript{\textregistered}~\cite{cst2024}. At the time of this study, the stage-B5 cell parameters closely matched the specifications foreseen for the Muon Collider Demonstrator cavity~\cite{accettura2025muon, rogers2023demonstrator}. 
For this reason, the stage-B5 cavity is used throughout this work as the primary representative case for detailed analysis, providing a scalable benchmark for the upcoming Muon Collider Demonstrator facility~\cite{schulte2025jacow}. All fourteen stages of the cooling channel were systematically analyzed, enabling the identification of potential weak points and performance-limiting factors throughout the entire 6D cooling sequence. Finally, Sec.~\ref{sec:5} offers the concluding remarks. The results of our investigation establish critical performance benchmarks and provide design guidelines for the RF systems of future muon collider rectilinear cooling channels.

\section{Cavity Design Methodology}\label{sec:2}

\begin{figure}[!htb]
   \centering
   \includegraphics*[width=0.9\columnwidth]{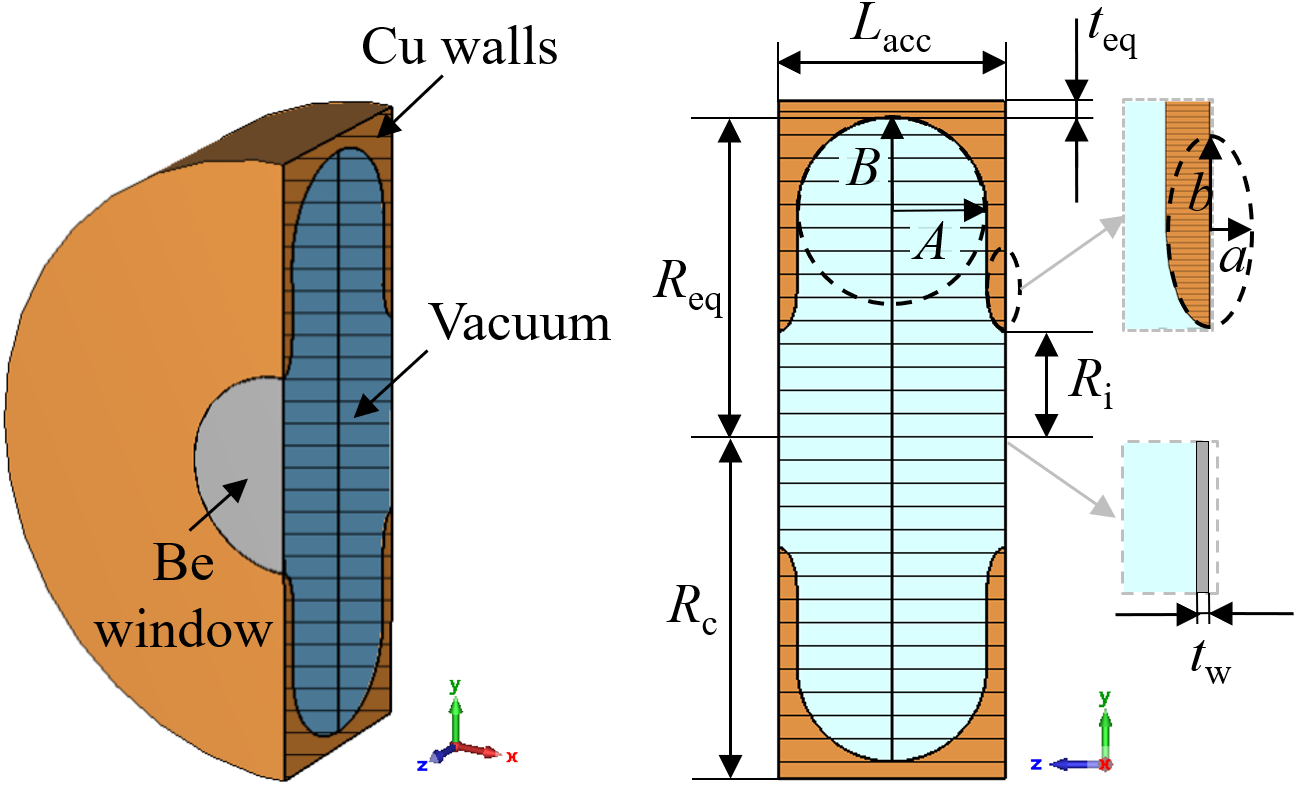}
   \vspace{0.1\baselineskip}
   \caption{Single-cell RF cavity design for muon cooling. The geometry is defined by the cavity accelerating length $L_{\mathrm{acc}}$, equatorial radius $R_{\mathrm{eq}}$, window radius $R_{\mathrm{i}}$, window thickness $t_{\mathrm{w}}$, and elliptical profile semi-axes $A, B$ and $a, b$. The thin Be windows seal both irises of the cell. Reproduced from~\cite{Barbagallo2024}.} 
   \label{fig:Single_cell_design}
\end{figure}

The RF cavities for each stage of the rectilinear cooling channel were designed to operate at the fundamental mode (FM) frequency $f_0$, based on the beam dynamics study~\cite{PhysRevAccelBeams.28.041003}. The resulting parameters for the single-cell design are reported in Table~\ref{tab:beam_dynamics_parameters}. The cavities are designed to accelerate muons in the $\mathrm{TM}_{010}$ mode, where the electric field is purely longitudinal, and the magnetic field vanishes at the cavity axis. They will operate in a pulsed mode, with a repetition rate $f_b$ of 5~Hz and a duty factor ($DF$) on the order of $10^{-4}$. The cavity cell geometry, initially presented in our previous work~\cite{Barbagallo2024}, follows the elliptical cavity parametrization described in~\cite{Shemelin2009}, as illustrated in Fig.~\ref{fig:Single_cell_design}.   

For each cooling stage, the single-cell cavity accelerating length $L_{\mathrm{acc}}$, the required accelerating gradient $E_{\mathrm{acc}}$, and the corresponding muon beam relativistic velocity factor $\beta=v/c$, where $v$ is the particle velocity and $c$ is the speed of light in vacuum, were adopted from~\cite{PhysRevAccelBeams.28.041003}. The single-cell RF cavity consists of Cu walls and thin Be windows closing both irises on either side of the cell (Fig.~\ref{fig:Single_cell_design}). Unlike conventional RF cavities, which employ open irises, closing the irises with metallic windows electromagnetically seals the cavity, preserving the on-axis electric field and maximizing the effective accelerating gradient. Indeed, a high gradient is essential in a muon cooling channel to rapidly restore longitudinal beam energy before significant particle decay occurs due to the short muon lifetime. Beryllium is selected as window material because its low atomic number ($Z=4$) minimizes beam degradation as muons pass from one cell to another. The window radius $R_{\mathrm{i}}$, equal to the cavity iris radius, and the thickness $t_{\mathrm{w}}$ were likewise specified in~\cite{PhysRevAccelBeams.28.041003} to ensure adequate beam transmission while minimizing particle scattering and energy loss in the window material. To respect practical engineering limits ($R_{\mathrm{i}} \lesssim 0.5 R_{\mathrm{eq}}$), the beam aperture radius $R_{\mathrm{i}}$ in Stages A1 and B1 was slightly adjusted prior to 3D optimization (reduced from $280\,\text{mm}$ to $240\,\text{mm}$ in Stage A1, and from $230\,\text{mm}$ to $210\,\text{mm}$ in Stage B1) compared to the initial baseline values of~\cite{PhysRevAccelBeams.28.041003}.

Since each cell is fully enclosed on both ends by windows, inter-cell RF coupling as in conventional multi-cell linacs cannot be implemented. Consequently, each cell must be powered individually via a dedicated Fundamental Power Coupler (FPC) port, following a distributed coupling scheme, such as that explored in~\cite{Luo:2024utb}, to fit within the tight radial space inside the surrounding high-field superconducting solenoids. For simplicity, FPC ports, auxiliary tuners, and vacuum ports were omitted from the present simulation model.

\begin{table*}
	\caption{Main input beam dynamics parameters for the design of the single-cell RF cavities in the rectilinear cooling channel from~\cite{PhysRevAccelBeams.28.041003}.}
    \setlength{\tabcolsep}{6pt}
		\begin{tabular}{cccccccccc}
          \toprule
			& $f_{0}$ & $L_\mathrm{acc}$ & $N_\mathrm{cell}$ & $R_\mathrm{i}$ &  $t_\mathrm{w}$ & $E_\mathrm{nom}$ & $\beta$ & $\sigma_{z}$ & $Q_{b}$ \\
			Stage & [MHz] & [mm] & [1] & [mm] &  [\textmu m] & [MV/m] & [1] & [mm] & [\textmu C] \\
			\hline
			A1 & 352 & 190 & 6 & 240 & 120 & 27.4 & 0.923 & 95.44 & 0.37    \\
			A2 & 352 & 190 & 4 & 160 & 70  & 26.4 & 0.894 & 59.51 & 0.28    \\
            A3 & 704 & 95  & 5 & 100 & 45  & 31.5 & 0.894 & 32.97 & 0.23     \\
            A4 & 704 & 95  & 4 & 80  & 40  & 31.7 & 0.901 & 22.41 & 0.19     \\
            B1 & 352 & 250 & 6 & 210 & 100 & 21.2 & 0.886 & 56.58 & 2.98     \\
            B2 & 352 & 220 & 5 & 190 & 80  & 21.7 & 0.885 & 57.05 & 2.54    \\
            B3 & 352 & 190 & 4 & 125 & 50  & 24.9 & 0.887 & 41.22 & 2.27     \\
            B4 & 352 & 220 & 3 & 95 &  45  & 24.3 & 0.886 & 36.13 & 1.98     \\
            B5 & 704 & 95  & 5 & 60 &  30  & 22.5 & 0.889 & 27.71 & 1.78     \\
            B6 & 704 & 95  & 4 & 45 &  20  & 28.0 & 0.888 & 23.64 & 1.59    \\
            B7 & 704 & 95  & 4 & 38 &  20  & 28.5 & 0.887 & 21.81 & 1.41     \\
            B8 & 704 & 95  & 4 & 28 &  20  & 27.1 & 0.884 & 20.97 & 1.31     \\
            B9 & 704 & 95  & 4 & 23 &  10  & 29.7 & 0.881 & 19.60 & 1.15     \\
            B10& 704 & 95  & 4 & 20 &  10  & 24.9 & 0.884 & 19.64 & 0.98     \\
		\toprule
        \end{tabular}
          
	\label{tab:beam_dynamics_parameters}
\end{table*}

Eigenmode simulations were performed using CST to optimize the cavity geometry, with the main focus on the cavity FM. The geometrical parameters $A = B$, which define the peak magnetic field region, and $a$ and $b$, which determine the cavity iris ellipse, were optimized primarily to reduce the peak surface electric field $E_{\mathrm{peak}}$, while maintaining a high cavity shunt impedance $(R/Q) \cdot Q_{0}$, where $(R/Q)$ is the geometric shunt impedance and $Q_{0}$ the intrinsic quality factor of the FM. The $(R/Q)$ is calculated by accounting for the transit-time factor $TTF$ as~\cite{wangler2008rf}

\begin{equation}
(R/Q)
= \frac{|V_z(0,0)|^{2}}{\omega_0U_0}TTF^2,
\label{eq:Geometric_shunt_impedance}
\end{equation}

\noindent
where $\omega_0$ and $U_0$ are the cavity angular frequency and the stored energy, respectively. Here, the on-axis longitudinal voltage, $V_z(0,0)$, is defined as
\begin{equation}
V_z(0,0) = \int_{z_{\mathrm{min}}}^{z_{\mathrm{max}}} E_z(0,0)\,dz,
\end{equation}
where $E_z(0,0)$ is the on-axis longitudinal electric field, while $z_{\mathrm{min}}$ and $z_{\mathrm{max}}$ define the integration boundaries along the cavity length. Note that Eq.~(1) uses the accelerator definition of $(R/Q)$, which is adopted consistently throughout this paper.

\begin{figure}[!htb]
   \centering
   \includegraphics*[width=0.9\columnwidth]{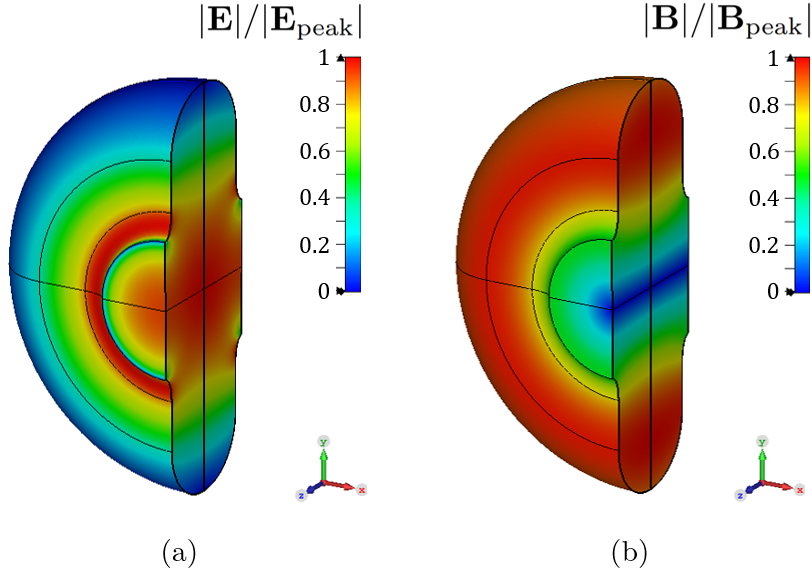}
   \vspace{0.1\baselineskip}
   \caption{Normalized electric (a) and magnetic field (b) distributions for the FM of the single-cell RF cavity with Be windows for muon cooling~\cite{Barbagallo2024}.}
   \label{fig:Field_single_cell}
\end{figure}

This optimization aims at reducing the surface losses $P_{\mathrm{diss}}$ on the window and cavity walls, given by 

\begin{equation} 
P_{\mathrm{diss}}=\frac{V_{\mathrm{acc}}^{2}}{(R/Q)\cdot Q_{0}},
\label{Peak dissipated power}
\end{equation}

\noindent
where $V_{\mathrm{acc}} = E_{\mathrm{acc}} L_{\mathrm{acc}}$ is the cavity accelerating voltage. The peak surface electric field $E_{\mathrm{peak}}$ was minimized to mitigate the risk of electron field emission. Firstly, with the cavity geometry tuned to the operating frequency, the ratio $r=b/a$ was swept at a fixed $a$ to identify the value of $r$ that minimizes $E_{\mathrm{peak}}$. Then, with the parameter $r$ fixed at the optimal value, the parameter $a$ was changed to further reduce $E_{\mathrm{peak}}$. At each step, the equator radius $R_{\mathrm{eq}}$ is adjusted to tune the cavity to its operating frequency. The outer cavity radius $R_{\mathrm{c}}$ is defined as $R_{\mathrm{c}}=R_{\mathrm{eq}}+t_{\mathrm{eq}}$, where $t_{\mathrm{eq}}=a$ represents the equatorial wall thickness. Ultimately, the parameter $a$ was fixed to 20~mm for the 352~MHz cavities and 10~mm for the 704~MHz ones, balancing $E_{\mathrm{peak}}$ minimization, an adequate $R/Q$, and thermal and mechanical stability. The normalized electric and magnetic field maps for the FM of the designed stage-B5 single-cell RF cavity are depicted in Fig.~\ref{fig:Field_single_cell}.

\begin{table*}
	\caption{Optimized geometrical parameters and RF figures of merit obtained from 3D CST electromagnetic simulations for the single-cell RF cavities in the rectilinear cooling channel.}
    \setlength{\tabcolsep}{1.9pt}
		\begin{tabular}{ccccccccccccccccccccc}
          \toprule
			& A & B & a & b & $R_{\mathrm{eq}}$ & $Q_{0}$ & $t_{\mathrm{f}}$ & $DF$ &  $TTF$  & $(R/Q)(\beta)$ & $P_\mathrm{diss}$ & $P_\mathrm{diss,Be}$ & $P_\mathrm{ave,Cu}$ &  $P_\mathrm{ave,Be}$ & $E_{\mathrm{peak,Cu}}$ & $E_{\mathrm{peak,Be}}$ \\
			Stage & [mm] & [mm] & [mm] & [mm] & [mm] & [$10^4$] & [\textmu s] &[$10^-4$] & [1] & [$\Omega$] & [MW] & [MW] & [W] & [W] & [MV/m] & [MV/m] \\
			\hline
			A1 & 75 & 75 & 20 & 44 & 359.08 & 3.05 & 31.203 & 1.17 & 0.906 &  171.73 & 4.24 & 1.60 & 310.82 & 187.92 & 11.72 & 27.38    \\
			A2 & 75 & 75 & 20 & 44 & 359.56 & 3.14  & 32.087 & 1.21 & 0.901 &  149.68  & 4.34 & 0.37 & 480.52 & 44.60 & 23.25 &  26.51   \\
            A3 & 37.5 & 37.5 & 10 & 22 & 180.82 & 2.20  & 11.248 & 4.27 & 0.901 & 160.36  & 2.06 & 0.41 & 70.11 & 17.68 & 20.80 &  31.51    \\
            A4 & 37.5 & 37.5 & 10 & 22 & 179.78 & 2.22  & 11.345 & 4.31 & 0.902 & 150.21  & 2.21 & 0.19 & 87.23 & 8.10 & 27.87 &  31.83   \\
            B1 & 105 & 105 & 20 & 44 & 362.19 & 3.91  & 39.954 & 1.50 & 0.828 & 183.70 & 2.68 & 0.62 & 310.29 & 92.61 &  12.16 &  21.25    \\
            B2 & 90 & 90 & 20 & 44 & 361.56 & 3.56  & 36.323 & 1.37 & 0.864 & 170.47  & 2.81 & 0.46 & 321.45 & 63.17 & 15.25 &  21.76  \\
            B3 & 75 & 75 & 20 & 44 & 355.36 & 3.15  & 32.148 & 1.21& 0.899 & 141.27  & 4.07 & 0.12 & 478.39 & 15.15 & 26.18 &  24.43    \\
            B4 & 90 & 90 & 20 & 44 & 352.75 & 3.59  &36.710 & 1.38 & 0.871 & 154.02   & 3.92 & 0.04 & 536.32 & 5.04 & 27.73 &  22.82    \\
            B5 & 37.5 & 37.5 & 10 & 22 & 177.32 & 2.23  & 11.366 & 4.33 & 0.901 & 140.85   & 1.18 & 0.03 & 49.83 & 1.32 & 24.12 & 22.03     \\
            B6 & 37.5 & 37.5 & 10 & 22 & 175.15 & 2.22  &11.360 & 4.31 & 0.902 & 137.40   & 1.89 & 0.01 & 80.60 & 0.60 & 33.29 & 26.51    \\
            B7 & 37.5 & 37.5 & 10 & 22 & 174.04 & 2.22  &11.354 & 4.31 & 0.904 & 136.87   & 1.97 & 0.01 & 84.55 & 0.26 & 35.25 &  25.98    \\
            B8 & 37.5 & 37.5 & 10 & 22 & 172.94 & 2.22  &11.347 & 4.31 & 0.907 & 137.39   & 1.79 & 1.5$\times10^{-3}$ & 76.85 & 0.06 & 34.67 &  22.89   \\
            B9 & 37.5 & 37.5 & 10 & 22 & 172.37 & 2.22  &11.344 & 4.31& 0.909 & 138.11   & 2.14 & 6.8$\times10^{-4}$ & 92.32 & 0.03 &  38.53 &   23.27   \\
            B10 & 37.5 & 37.5 & 10 & 22 & 172.23 &2.22 & 11.341 & 4.31 & 0.912 & 139.03   & 1.51 & 2.3$\times10^{-4}$ & 64.87 & 0.01 & 32.51 &  18.34    \\
		\toprule
        \end{tabular}
          
	\label{tab:RF_parameters}
\end{table*}

\subsection{RF performance of optimized cavities}
\label{twdwp}

Table~\ref{tab:RF_parameters} summarizes the main optimized geometrical parameters and RF figures of merit for the FM, obtained from CST simulations, for the single-cell RF cavities of the rectilinear cooling channel. The calculated values for the intrinsic quality factor $Q_{0}$ and the corresponding filling time $t_{\mathrm{f}}$ are consistent with typical values for normal-conducting cavities at the considered operating frequencies~\cite{wangler2008rf, grudiev2021muon}. Here, the filling time represents the duration required to fill the cavity to the nominal voltage, $V_{\mathrm{nom}} = E_{\mathrm{nom}} L_{\mathrm{acc}}$  (where $E_{\mathrm{nom}} = E_{\mathrm{acc}} / TTF$ is the nominal gradient), and is defined as~\cite{wangler2008rf}
\begin{equation}
t_{\mathrm{f}} \approx \frac{2Q_{\mathrm{L}}}{\omega_{0}} \ln\left(\frac{2\beta_{\mathrm{c}}}{\beta_{\mathrm{c}} - 1}\right),
\label{eq:filling_time}
\end{equation}
where $Q_{\mathrm{L}}=Q_{0}/(1+\beta_{\mathrm{c}})$ is the loaded quality factor and $\beta_{\mathrm{c}}$ is the cavity coupling factor. Our analysis considers a fixed coupling factor of $\beta_{\mathrm{c}}=1.2$, as this value was identified as a reasonable compromise between reducing the filling time, and thus the average power dissipation, and limiting the peak power required from the generator~\cite{Barbagallo:2024}. The cavity voltage profile during cavity filling is given by~\cite{Barbagallo2024}
\begin{equation}
V_{\mathrm{f}}(t) = V_{\mathrm{nom}} \frac{2\beta_{\mathrm{c}}}{1+\beta_{\mathrm{c}}}\left(1 - \mathrm{e}^{-\omega_{0} t/2Q_{\mathrm{L}}}\right).
\label{eq:filling_decay_voltage}
\end{equation}

\noindent
Following the duration of the injected bunch train at the nominal voltage ($t_{\mathrm{b}} = \SI{0.1}{\micro\second}$), the cavity voltage decays according to $V_{\mathrm{d}}(t)=V_{\mathrm{nom}} \mathrm{e}^{-\omega_{0} t/2Q_{\mathrm{L}}}$. The cavity duty factor $DF$ is then calculated as the ratio between the average power and the peak dissipated power: 

\begin{equation}
     DF = \frac{P_{\mathrm{ave}}}{P_{\mathrm{diss}}} = \frac{\int_{0}^{\infty}P(t)\mathrm{d}t}{V_{\mathrm{acc}}^{2}/(R/Q \cdot Q_{0})},
\label{eq:duty_factor} 
\end{equation}
\noindent
where $P(t)$ is the time-dependent power calculated from the cavity voltage profile and $V_{\mathrm{acc}}=E_{\mathrm{acc}}L_{\mathrm{acc}}$ the accelerating cavity voltage. The short cavity length in all stages ensures a high transit-time factor, maintaining excellent phase-velocity matching between the RF field and the beam. The geometrical shunt impedance $R/Q$ values remain relatively stable across the different cavity stages. Although $R/Q$-values are lower than those of an ideal pillbox cavity (for example, the $R/Q$ of the stage-B5 cavity exhibits a 19\% reduction compared to a pillbox at 704~MHz), the achieved values represent the maximum attainable ones within the geometrical constraints imposed by the beam dynamics design. In the pre-merging section, the 352~MHz cavities exhibit higher power dissipation compared to the 704~MHz cavities, primarily due to their larger dimensions. A similar trend is observed in the post-merging section. Only a small fraction of the total power $P_\mathrm{diss}=P_\mathrm{diss,Cu}+P_\mathrm{diss,Be}$ is dissipated on the Be-window surface, denoted as $P_\mathrm{diss,Be}$. Using the calculated $DF$, the average dissipated power values for the copper walls ($P_{\mathrm{ave,Cu}} = DF \cdot P_{\mathrm{diss,Cu}}$) and Be windows ($P_{\mathrm{ave,Be}} = DF \cdot P_{\mathrm{diss,Be}}$) are evaluated. This dissipation on the windows increases with their radius, as larger radii extend further into the high-magnetic-field region of the cavity. Nevertheless, both peak and average dissipated power on the cavity walls remain within acceptable limits for copper operation. Multiplying the cavity length, the geometrical shunt impedance $R/Q$, and power values by the number of cavity cells $N_{\mathrm{cell}}$ yields the equivalent values for the multi-cell RF cavity of the considered stage.

Table~\ref{tab:RF_power_requirements} lists the RF power requirements for the single-cell cavities at each stage of the rectilinear cooling channel. The total peak power per cavity cell is obtained as $P_{\mathrm{g}}=P_{\mathrm{diss}}\beta_{\mathrm{c}}$, where $\beta_{\mathrm{c}}$ is the coupling factor. The total peak RF power per stage, $P_{g,\text{tot}}$, is obtained by multiplying the peak power per cavity cell by the total number of single-cell RF cavities in that cooling stage, $N_{\mathrm{tot}}$.

\begin{table}
	\caption{Single-cell RF power requirements and stage-level power scaling for the rectilinear cooling channel. $N_{\mathrm{tot}}$ represents the total number of single cells in the stage ($N_{\mathrm{tot}} = N_{\mathrm{cav}} \cdot N_{\mathrm{cell}}$).}
    \setlength{\tabcolsep}{3.5pt}
		\begin{tabular}{ccccccc}
          \toprule
			& $P_{\mathrm{g}}$ & $DF_\mathrm{g}$ & $N_\mathrm{tot}$ &  $P_\mathrm{g,tot}$ & $P_\mathrm{g,ave,tot}$ & $P_\mathrm{plug,tot}$  \\
			Stage & [MW] & [$10^{-4}$] & [-] & [MW] & [kW] & [kW] \\
			\hline
			A1 & 5.09 & 1.56 & 348 & 1772.7 & 277.1 & 439.8 \\
			A2 & 5.21 & 1.61 & 356 & 1854.9 & 297.9 & 472.8 \\
            A3 & 2.47 & 0.57 & 405 & 999.4 & 56.7 &  90.0 \\
            A4 & 2.66 & 0.57 & 496 & 1317.1 & 75.4 & 119.7 \\
            B1 & 3.21 & 2.08 & 132 & 424.2 & 88.1 & 139.8 \\
            B2 & 3.37 & 2.10 & 185 & 623.1 & 130.7 & 207.4 \\
            B3 & 4.88 & 1.61 & 240 & 1171.6 & 188.8 & 299.7 \\
            B4 & 4.70 & 1.84 & 165 & 775.7 & 142.9 & 226.9 \\
            B5 & 1.42 & 0.57 & 275 & 390.1 & 22.4 & 35.51 \\
            B6 & 2.26 & 0.71 & 220 & 497.7 & 35.3 & 56.1 \\
            B7 & 2.36 & 0.61 & 160 & 378.0 & 23.2 & 36.8 \\
            B8 & 2.14 & 0.62 & 284 & 608.5 & 37.5 & 59.4 \\
            B9 & 2.57 & 0.57 & 208 & 535.1 & 30.6 & 48.5 \\
            B10& 1.81 & 0.57 & 188 & 339.8 & 19.4 & 30.9 \\
            \hline
            Total &  &  &  & 11687.9 & 1425.9 & 2263.3 \\
		\toprule
        \end{tabular}
          
	\label{tab:RF_power_requirements}
\end{table}

Consequently, the number of multi-cell cavities per stage is given by $N_{\mathrm{cav}}=N_{\mathrm{tot}}/N_{\mathrm{cell}}$. The average RF power per stage, $P_{\text{g,ave,tot}}$, accounts for the computed generator duty factor $DF_{\mathrm{g}}$, which is defined as the ratio between the average power of the generator for a single-cavity cell and its peak input RF power:

\begin{equation}
    DF_{\mathrm{g}} = \frac{P_{\mathrm{g,ave{}}}}{P_{\mathrm{g}}} = \frac{P_{\mathrm{g}} t_{\mathrm{f}} \cdot f_{\mathrm{b}}}{P_{\mathrm{g}}}.
\label{eq:duty_factor_generator} 
\end{equation}

\noindent
Note that in this operational regime, because the stored electromagnetic energy continues to dissipate in the cavity walls after the generator pulse ends ($t > t_{\mathrm{f}}$), the exponential decay tail contributes significantly to the time-integrated losses, resulting in $DF > DF_{\mathrm{g}}$.
The total wall-plug power for the RF system of each stage is then evaluated as
\begin{equation}
    P_{\text{plug,tot}} = \frac{P_{\text{g,ave,tot}}}{\eta_{\mathrm{g}}\eta_{\mathrm{m}}},
\label{eq:plug_power}
\end{equation}
assuming a generator efficiency of $\eta_{\mathrm{g}} = 0.7$ and a modulator efficiency of $\eta_{\mathrm{m}} = 0.9$. Overall, the total power consumption for the RF system remains within acceptable limits for pulsed normal-conducting RF operation at the considered duty factors.

\subsection{Comparison of cavities with and without windows}
\label{cavity_comparison}

The use of Be windows in the RF cavities for the rectilinear cooling channel presents both performance advantages and limitations. The windows preserve the on-axis accelerating field, maintaining cavity performance close to that of an ideal pillbox. This results in an increased integrated longitudinal voltage on the cavity axis and, consequently, a higher $R/Q$, which in turn reduces the RF power dissipated on the cavity and window walls for a given accelerating gradient. Moreover, the peak surface electric field at the irises is lowered. Consequently, this design choice directly enhances muon acceleration while reducing the risk of RF breakdown. However, the use of very thin Be windows introduces several technical and operational challenges. In the current design, the window thickness varies across the channel from~\SI{120}{\micro\meter} in stage A1 down to~\SI{10}{\micro\meter} in stage B10 (see Table~\ref{tab:RF_parameters}). This progressive reduction in thickness is strictly dictated by beam dynamics optimization~\cite{PhysRevAccelBeams.28.041003} to minimize multiple Coulomb scattering, thereby preventing transverse emittance growth and maximizing beam transmission.

The thermo-mechanical behavior of thin beryllium windows in RF cavities for muon cooling has been extensively investigated in prior multiphysics studies~\cite{Li:1998linac, Li:EPAC02, li2003201}.  Even though reducing beryllium window thickness increases the center-to-edge temperature rise ($\Delta T\propto~\!\!1/t_\mathrm{w}$)~\cite{Li:1998linac}, this effect is heavily counteracted by the steep fourth-power drop in dissipated RF power as the aperture radius decreases ($P_{\mathrm{diss,Be}} \propto R_{\mathrm{i}}^4$). Consequently, even for ultrathin windows down to \SI{10}{\micro\meter} in final stages of the rectilinear cooling channel, the dissipated power on the windows drops to a negligible fraction of the total cavity loss, resulting in minimal thermal load. From a mechanical perspective, reducing the window radius lowers structural stresses and deformations, effectively compensating for the reduced thickness in the last cooling stages. However, mechanical deformation resulting from
Lorentz-force radiation pressure, which scales with the square of the accelerating gradient, can lead to cavity frequency detuning~\cite{li2003201, Li:EPAC02, Barbagallo2024}. Furthermore, pulsed-beam operation may induce structural fatigue of such thin windows, posing a non-negligible risk of rupture. In the event of window failure, cavity contamination may occur, potentially compromising RF performance. Thermal deformations of the cavity windows were found to be negligible under optimal water cooling conditions, in both steady-state and transient regimes~\cite{Barbagallo2024, Boff_Indico_2025}. Adopting a curved window geometry, like the one considered for the muon cooling cell design~\cite{li2003201} and in the more recent design under development~\cite{Boff_Indico_2025}, can improve window stiffness. This geometric feature mitigates the total window deformation and the resulting cavity frequency detuning, without degrading RF performance. 

\begin{figure}[!htb]
   \centering
   \includegraphics*[width=0.7\columnwidth]{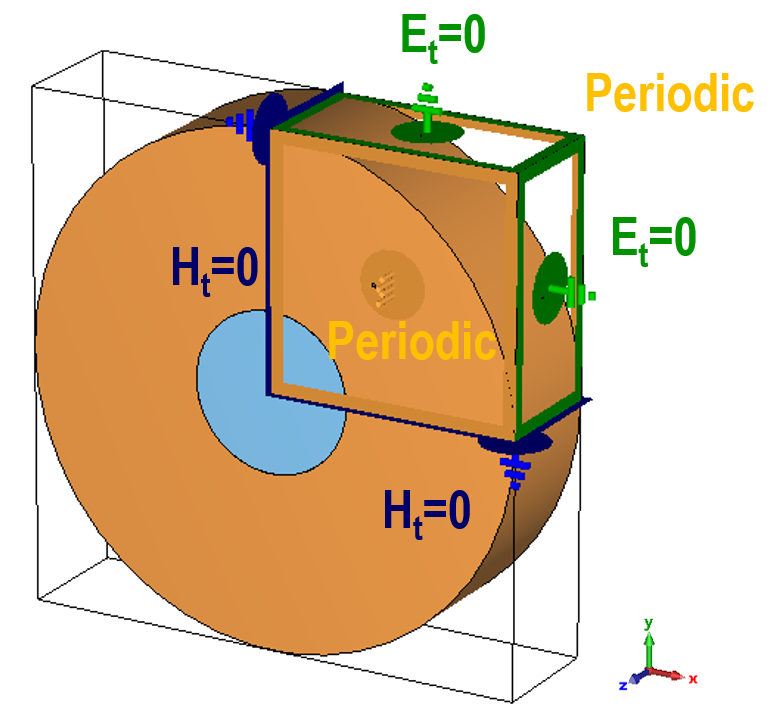}
   \vspace{0.1\baselineskip}
   \caption{Single-cell simulation setup with periodic boundary conditions and symmetry planes, used to model the field distribution of a multi-cell cavity without Be windows. The cavity utilizes the same elliptical profile as the stage B5 in Fig.~\ref{fig:Single_cell_design}, with its equatorial radius $R_{\mathrm{eq}}$ re-tuned to maintain a resonant frequency of 704 MHz. Reproduced from~\cite{Barbagallo:2025}.}
   \label{fig:Periodic_boundaries}
\end{figure}

For comparison, we simulated an equivalent open-iris single-cell cavity without Be windows. We used the same elliptical profile as stage B5, with the equatorial radius $R_{\mathrm{eq}}$ re-tuned to maintain a resonant frequency of 704~MHz. The single-cell simulation setup, implemented in CST, is illustrated in Fig.~\ref{fig:Periodic_boundaries}. The simulation employed periodic boundary conditions along the beam axis (the $z$-direction), with an RF phase advance of $\phi$ between adjacent cells to model the field distribution of a multi-cell structure. This RF phase advance is defined as $\phi = ({2 \pi L_\mathrm{acc}})/({\lambda_{0} \, \beta})$, where $\lambda_{0}=c/f_{0}$ is the wavelength at the FM frequency. Perfect electric conductor (PEC) boundary conditions (i.e., $\mathbf{E}_t = 0$) were applied to the transverse boundaries. Perfect magnetic conductor (PMC) boundary conditions (i.e., $\mathbf{H}_t = 0$) were applied on the $YZ$ and $XZ$ symmetry planes to excite the FM field pattern and reduce the simulation domain. 

It is worth noting that the implemented setup with longitudinal periodic boundaries effectively models a single cell as part of an infinite multi-cell traveling-wave structure, which is operated under synchronous conditions, where the RF phase velocity is matched to the beam velocity. In this configuration, the reduction in the $TTF$ is less degraded than in a finite single-cell cavity closed by Be windows, which behaves as a finite standing-wave structure. As a consequence, higher values of $TTF$ can be obtained in the periodic model. This design yields a higher $TTF$,
resulting in a larger effective accelerating voltage experienced by the beam for a given on-axis field. The geometric shunt impedance per cavity cell is significantly reduced, resulting in higher peak power requirements. Moreover, the peak surface electric field at the cavity iris is higher than the design with windows, which increases the probability of RF breakdown in the cavity.

Table~\ref{tab:BeWindowComparison} compares the RF performance of the stage-B5 cavity cell with and without Be windows. 

\begin{table}[h]
\centering
\caption{Comparison of RF parameters for the stage-B5 cavity cell with and without Be windows.}
\label{tab:BeWindowComparison}
\begin{tabular}{lccc}
\toprule
Parameter & Unit & With Be window & Without Be window \\
\hline
$f_0$ & MHz & 704 & 704 \\
$Q_0$ & [1] & $2.23\times10^{4}$ & $2.25\times10^{4}$ \\
$TTF$ & [1]  & 0.90 & 0.99 \\
$(R/Q)(\beta)$ & $\Omega$ & 140.85 & 79.08 \\
$E_{\mathrm{acc}}$ & MV/m & 20.26 & 22.42 \\
$P_{\mathrm{diss}}$ & MW & 1.18 & 2.55 \\
$E_{\mathrm{peak}}$ & MV/m & 24.12 & 48.81 \\
\toprule
\end{tabular}
\end{table}
\noindent
While removing the Be windows from the investigated cavity slightly increases the $TTF$, the peak surface electric field nearly reaches 50~MV/m, and the dissipated power more than doubles.  
Given the similar geometry and operating parameters of the other cavity stages, comparable trends are expected throughout the rectilinear cooling channel. From this analysis, the small gain in $TTF$ does not justify removing the Be windows, as the main concerns remain power consumption and peak surface electric fields. This conclusion, however, will be revisited after the analysis of the wakefields and beam-loading effects in the next section.

\section{Wakefield analysis}\label{sec:3}
\label{wakefield_section}

A charged particle beam traversing an accelerating RF cavity excites electromagnetic fields, known as wakefields. These fields arise from the beam-cavity interaction and act back on the beam itself, potentially inducing beam-loading effects that reduce the accelerating voltage and degrade the beam quality~\cite{Weiland1992}. In standard 3D wakefield solvers, such as the Wakefield Solver in CST and ECHO3D~\cite{Zagorodnov2005}, the wake potential is computed considering an exciting bunch of Gaussian profile traversing the cavity. The resulting wakefield intrinsically contains both the resonant-mode excitations and space-charge contributions generated by the beam-cavity interaction.

For ultrarelativistic beams ($\beta\rightarrow1$), space-charge effects can be neglected. In the considered rectilinear muon cooling channel, the beam relativistic factor is in the range of $\beta\approx0.88-0.92$, and therefore space-charge effects become relevant. In this intermediate relativistic regime, separating cavity-driven wakefields from space-charge wakefields is essential for developing appropriate compensation techniques. Existing wakefield solvers do not provide a built-in approach to separate these contributions.
 
Our approach consists of a hybrid frequency-time-domain numerical method to isolate the space-charge contribution from the total wake potential. Specifically, the total longitudinal wake potential $W_{z,\mathrm{tot}}(t)$ can be decomposed as

\begin{equation}
W_{z,\mathrm{tot}}(t) = \underbrace{W_{z,\mathrm{FM}}(t) + W_{z,\mathrm{HOMs}}(t)}_{W_{z,\mathrm{eig}}(t)} + W_{z,\mathrm{SC}}(t),
\label{eq:Wtot}
\end{equation}

\noindent
where $W_{z,\mathrm{eig}}(t)$ represents the eigenmode wake potential, comprising the contributions from the fundamental mode (FM), $W_{z,\mathrm{FM}}(t)$, and higher-order modes (HOMs), $W_{z,\mathrm{HOMs}}(t)$, while $W_{z,\mathrm{SC}}(t)$ denotes the space-charge (SC) component. Here, the wake potential is shown as a function of time $t$, with $t=0$ defined at the center of the source bunch. This relates to the longitudinal position $z = \beta c t$ of the test particle, where $c$ is the speed of light in vacuum. The total wake potential is computed using the CST wakefield solver with a single Gaussian bunch. The single-charge wake function due to FM and HOMs can be reconstructed from the cavity mode spectrum and figures of merit obtained in the CST eigenmode solver, according to the following equation~\cite{Wilson1989}

\begin{equation}
W_{z,\mathrm{eig}}^{0}(t)
= H(t)\sum_{n}\frac{\omega_n}{2}(R/Q)_{n}\,e^{-t/\tau_n}\cos(\omega_n t),
\label{eq:Ws_total}
\end{equation}
\noindent
where \(H(t)\) is the Heaviside step function ensuring causality, $n$ is the mode index, \(\omega_n\) is the angular frequency of mode \(n\), $(R/Q)_{n}$ its geometric shunt impedance, and $\tau_n=2Q_{\mathrm{L},n}/\omega_n$ the field decay constant, with $Q_{\mathrm{L},n}$ denoting the loaded quality factor.

The eigenmode wake potential \(W_{z,\mathrm{eig}}(t)\) in Eq.~\eqref{eq:Wtot}, which is cavity-dependent, is then obtained through convolution of \(W_{z,\mathrm{eig}}^{0}(t)\) with the bunch charge distribution \(\lambda(t)\):

\begin{equation}
W_{z,\mathrm{eig}}(t)
= \frac{1}{Q_b} \int_{-\infty}^{t} W_{z,\mathrm{eig}}^{0}(t-t')\, \lambda(t')\, dt',
\label{eq:convolution}
\end{equation}

\noindent

\noindent
where the bunch charge distribution is normalized such that

\begin{equation}
\int_{-\infty}^{\infty} \lambda(t')\, dt' = Q_b,
\label{eq:normalization}
\end{equation}

\noindent
with $Q_b$ being the bunch charge. Specifically, for a Gaussian bunch of rms duration \(\sigma_{t}\), the charge distribution is

\begin{equation}
\lambda(t) = \frac{Q_b}{\sqrt{2\pi}\sigma_{t}} \exp\!\Big(-\frac{t^2}{2\sigma_{t}^2}\Big).
\label{eq:lambda_gaussian}
\end{equation}

The bunch charge varies across the different cooling stages due to particle transmission losses. The single-bunch charge, $Q_{b,i}$, listed for each stage in Table~\ref{tab:beam_dynamics_parameters}, is calculated from the bunch intensity, i.e., the number of muons per bunch, $N_{\mu,b,i}$, at that stage, as follows. At the end of the post-merging section of the rectilinear cooling channel, the bunch intensity is $N_{\mu,b} = 5.32 \times 10^{12}$~\cite{taylor2025mucol}. For both the pre- and post-merging sections, the intensity at the start of each stage $i$ is obtained from the final bunch intensity by accounting for the transmission efficiencies $\eta_{\mathrm{tr},j}$, defined as the ratio of the number of particles exiting stage $j$ to those entering it, for all downstream stages~\cite{accettura2025muon}
\begin{equation}
N_{\mu,b,i} = \frac{N_{\mu,b}}{\prod_{j=i}^{N_\mathrm{stages}} \eta_{\mathrm{tr},j}},
\end{equation}
\noindent
where $N_\mathrm{stages}$ is the total number of stages in the rectilinear cooling channel. The bunch charge at each stage is then
\begin{equation}
Q_{b,i} = N_{\mu,b,i} \cdot e,
\end{equation}
where $e = 1.602 \times 10^{-19}\,\mathrm{C}$ is the muon charge. The rms longitudinal bunch lengths, $\sigma_z$  (where $\sigma_t = \sigma_z / (\beta c)$), were obtained from beam-dynamics studies performed by R.~Zhu~\cite{ZhuPriv}.

{\subsection{Space-charge wake potential decomposition}}

The analytical formulation given in Eq.~\eqref{eq:Ws_total} does not account for the space-charge effects, which dominate the short-range wakefield regime. Therefore, according to Eq.~\eqref{eq:Wtot}, the residual term \(W_{{\color{blue} z,}\mathrm{SC}}(t)\) can be obtained by subtracting \(W_{{\color{blue} z,}\mathrm{eig}}(t)\) from the total wake potential \(W_{{\color{blue} z,}\mathrm{tot}}(t)\) computed in the wakefield solver. 

The longitudinal space-charge wake potential can be further decomposed into the sum of indirect (ISC) and direct (DSC) space-charge components:

\begin{equation}
W_{z,\mathrm{SC}}(t)
= W_{z,\mathrm{ISC}}(t) + W_{z,\mathrm{DSC}}(t).
\label{eq:Wsc}
\end{equation}
\noindent
The indirect space-charge wake potential $W_{z,\mathrm{ISC}}(t)$ is generated by the fields reflected by the cavity boundaries, which subsequently propagate back and interact with the beam along the cavity axis~\cite{Schindl1999}. The induced effect is therefore cavity-dependent. In contrast, the direct space-charge wake potential $W_{z,\mathrm{DSC}}(t)$ arises from the Coulomb field of the bunch itself and is weakly dependent on the cavity geometry and, to the first order, can be approximated as geometry-independent when the transverse dimensions of the structure are large compared to the transverse offset of the particle.

\begin{figure}[hbt!]
    \centering

    \begin{subfigure}{0.453\textwidth}
       \hspace*{-2em} 
        \centering
        \includegraphics[width=\textwidth]{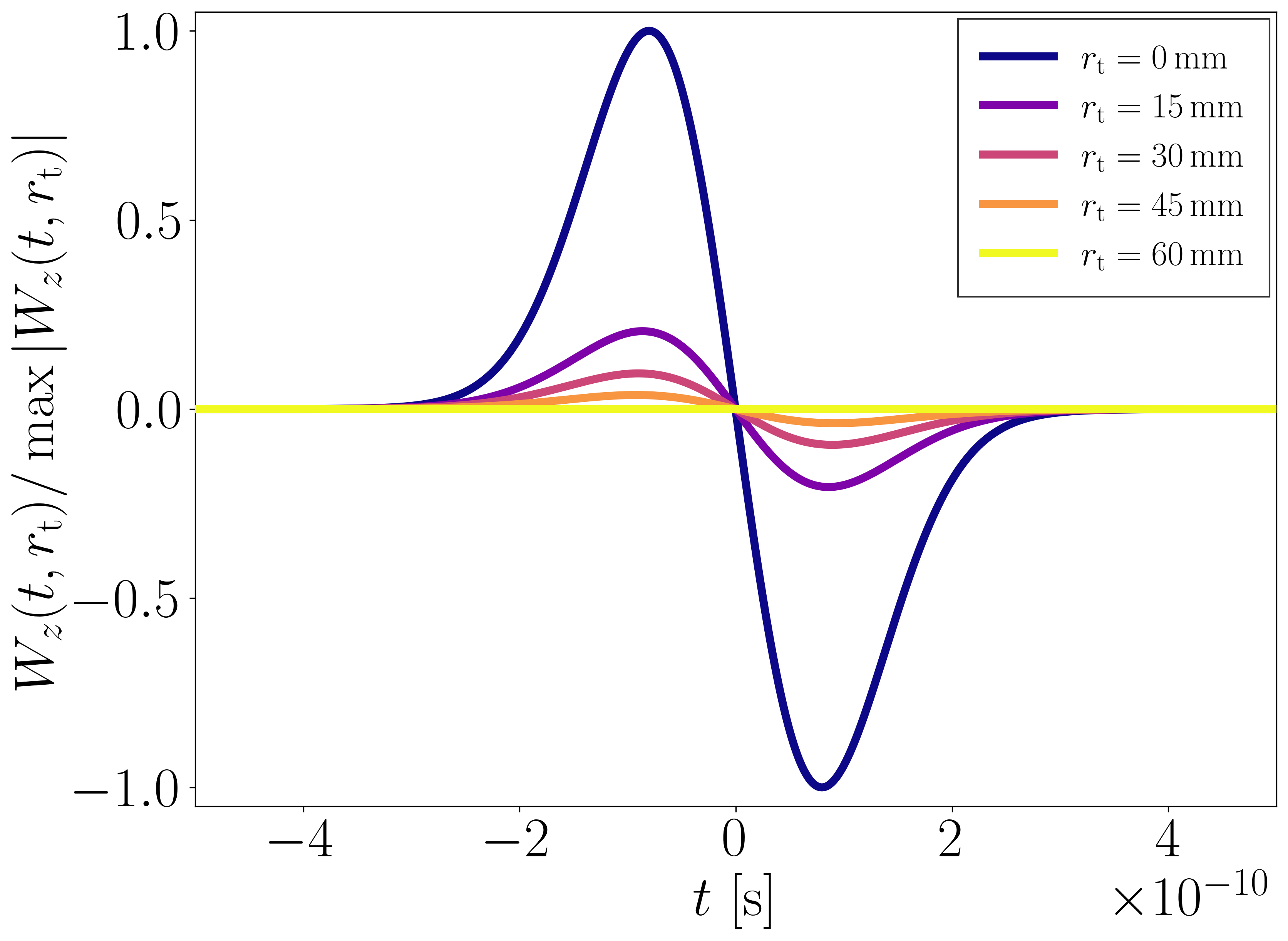}
        \caption{}
        \label{fig:Wz_raw}
    \end{subfigure}

    \vspace{0.5cm}

    \begin{subfigure}{0.453\textwidth}
        \centering
        \includegraphics[width=\textwidth]{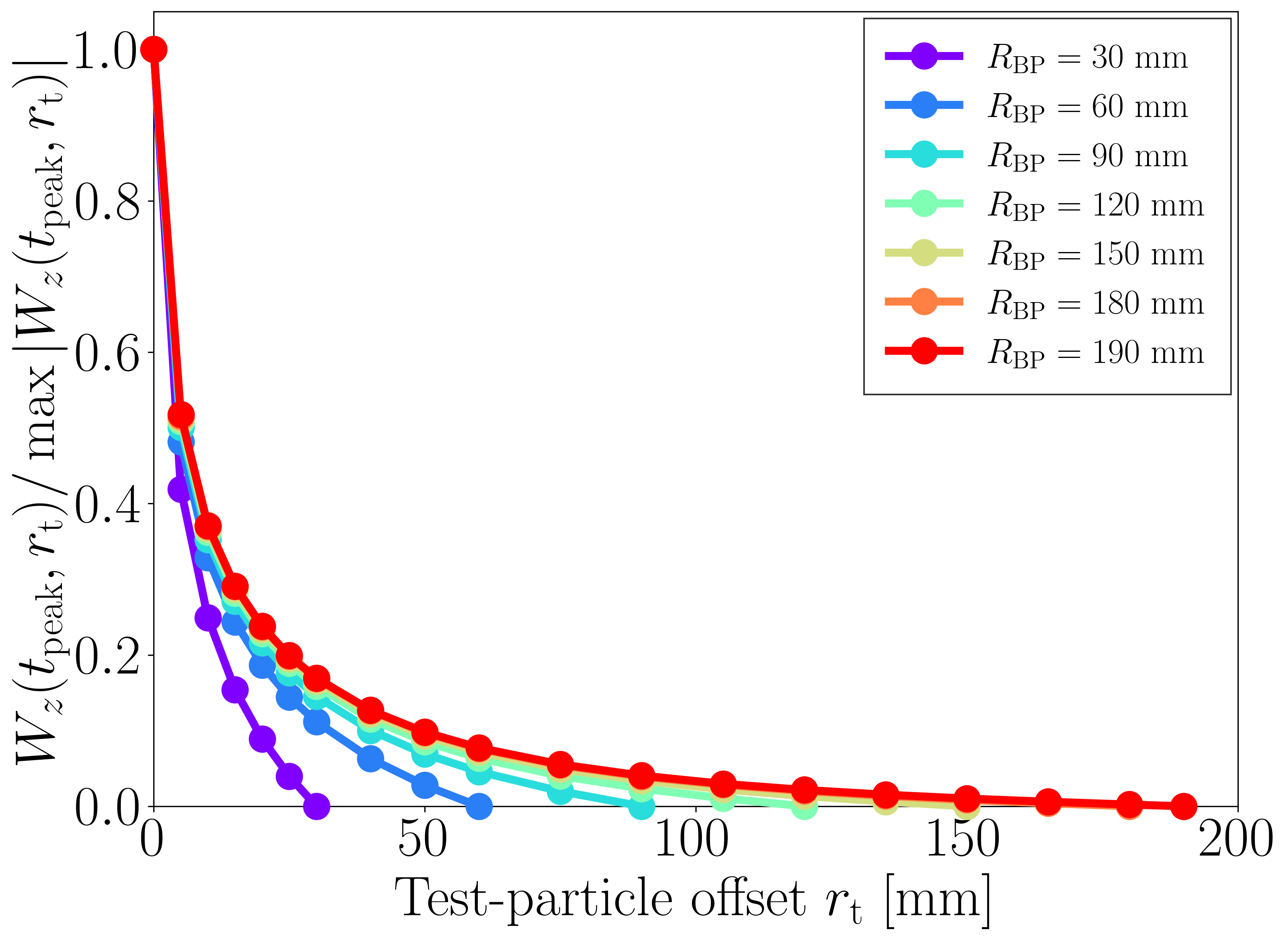}
        \caption{}
        \label{fig:Wz_norm}
    \end{subfigure}

    \caption{Longitudinal BP space-charge wake potential: (a) normalized wake for a BP of radius 60 mm at various test-particle transverse offsets; (b) normalized transverse dependence of the wake potential for several BP radii $R_\mathrm{BP}$.}
    \label{fig:Wake_BP}
\end{figure}

In the following, we propose a method to separate and quantify the ISC and DSC wake potentials in RF cavities. This procedure allows isolating cavity-dependent and independent components of the total wake potential for beam-loading compensation. To perform this decomposition, we first simulated a simplified model in the wakefield solver. This setup consists of a vacuum beam pipe (BP) of radius $R_{\mathrm{BP}}$ with PEC walls and open boundary conditions at its apertures, traversed by a subrelativistic Gaussian beam. In this configuration, cavity modes are not excited, and the computed wakefield contains only space-charge effects. The corresponding longitudinal space-charge wake potential induced by a source particle and acting on a test particle with transverse offset $r_{\mathrm{t}}$ can be written as

\begin{equation}
W_{z,\mathrm{SC}}^{\mathrm{BP}}(t,r_{\mathrm{t}},R_{\mathrm{BP}})
= W_{z,\mathrm{ISC}}^{\mathrm{BP}}(t,R_{\mathrm{BP}})+W_{z,\mathrm{DSC}}^{\mathrm{BP}}(t,r_{\mathrm{t}}).
\label{eq:Wsc_BP}
\end{equation}
\noindent
Here, the ISC contribution is a function of the beam pipe radius, while the DSC depends only on the transverse offset of the test charge. The behavior of these two components is illustrated in Fig.~\ref{fig:Wake_BP}.

The amplitude of the total longitudinal SC wake potential decreases rapidly with increasing transverse offset $r_{\mathrm{t}}$ (see Fig.~\ref{fig:Wz_raw}). This trend represents the DSC effect, which exhibits an approximate 1/$r_{\mathrm{t}}^{2}$ dependence, consistent with the scaling of the Coulomb field in free space~\cite{Schindl1999}. The monotonic 1/$r_{\mathrm{t}}^{2}$-like decay due to Coulomb interaction is clearer in Fig.~\ref{fig:Wz_norm}. This figure illustrates the wake potential values at $t_{\mathrm{peak}}$, defined as the time at which the wake reaches its maximum, normalized by the maximum value across the different $r_{\mathrm{t}}$ for different BP radii. Here, the vertical shift between curves corresponds to the ISC component. For small radii, the conducting boundaries strongly confine the induced image-charge fields, enhancing the ISC term. For large radii, BP walls are farther away from the beam axis, and the ISC contribution vanishes, i.e., $W_{z,\mathrm{SC}}^{\mathrm{BP}}(t,r_{\mathrm{t}},R_{\mathrm{BP}})~\approx~W_{z,\mathrm{DSC}}^{\mathrm{BP}}(t,r_{\mathrm{t}})$.

This observation allows us to approximate the DSC wake potential component in a cavity whose equator radius $R_{\mathrm{eq}}$ is much larger than the transverse offset of the test particle $r_{\mathrm{t}}$. In these wakefield simulations, the beam is assumed to have zero transverse size, so the validity of this approximation depends only on the distance between the beam axis and the cavity boundaries. In this limit, boundary-induced fields vanish, and 
the direct space-charge wake potential of the cavity in Eq.~\eqref{eq:Wsc} can be approximated by the direct space-charge contribution of the reference beam-pipe model as 
\begin{equation}
W_{z,\mathrm{DSC}}(t,r_{\mathrm{t}},R_{\mathrm{eq}})~\approx~W_{z,\mathrm{DSC}}^{\mathrm{BP}}(t,r_{\mathrm{t}}), \;\;\;  R_{\mathrm{eq}} \gg r_{\mathrm{t}}.
\label{eq:WDSC_cavity}
\end{equation}

\vspace{0.8em}

{\subsection{Wake potential decomposition results}}
\label{subsec:decomposition_results}⁠

\begin{figure}[hbt!]
    \includegraphics[width=0.48\textwidth]{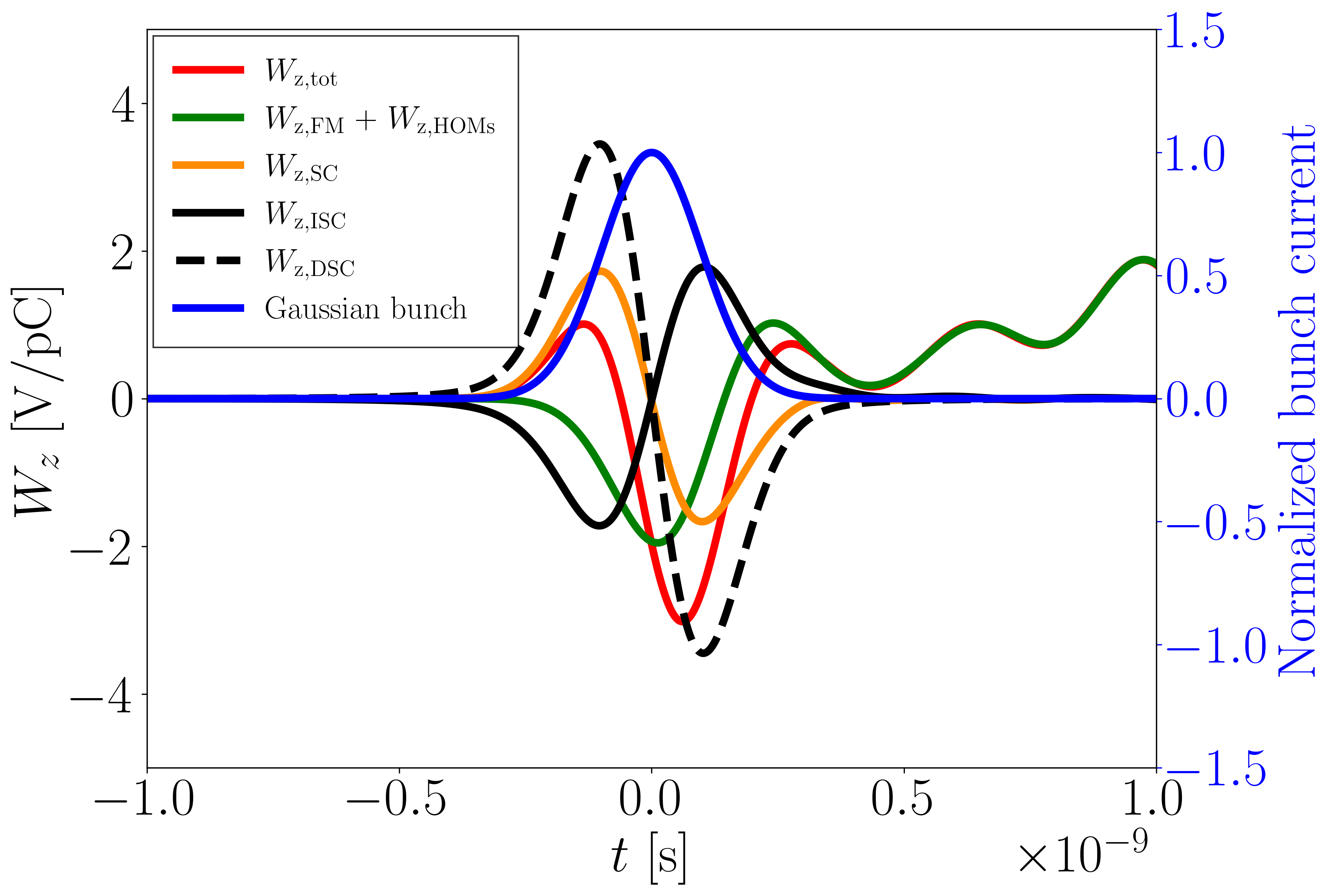}
    \caption{Longitudinal wake potential decomposition for the stage-B5 multi-cell cavity with Be windows. The reconstructed eigenmode wake potential includes the first nine monopole modes up to 4~GHz. The normalized Gaussian bunch longitudinal profile is shown in the secondary axis.}
    \label{fig:Wake_dec_B5_multi_cell}
\end{figure}

Figure~\ref{fig:Wake_dec_B5_multi_cell} shows the computed components of the wake potential for the stage-B5 multi-cell cavity with Be windows, together with the normalized Gaussian bunch used in the simulations. The reconstructed eigenmode wake potential accounts only for transverse magnetic (TM) monopole modes up to a cutoff frequency beyond which the bunch spectral content becomes negligible, i.e., above the $-20~\mathrm{dB}$ spectral field amplitude level, which corresponds to $f_{-20\mathrm{dB}} \approx 3.44~\mathrm{GHz}$ for the stage B5 cavity. Including modes up to a slightly higher frequency, such as $4.0~\mathrm{GHz}$ in this case, provides a conservative margin. The same mode truncation criterion is applied to the cavities across all other stages in the cooling channel. Dipole, quadrupole, and modes of higher order are excluded, as they exhibit no induced voltage on the axis~\cite{Sangho2002}. 

Table~\ref{tab:eigenmodes} lists the modes of the stage-B5 single-cell cavity with Be windows used in the wake potential reconstruction, along with their corresponding parameters. Multiplying the $R/Q$-values by $N_{\mathrm{cell}}$ yields the equivalent values for a multi-cell RF cavity. The total wake potential is calculated using a Gaussian beam with longitudinal bunch length $\sigma_{z}$ and a total charge $Q_b$, exciting a wake length of $100\sigma_{z}$. 

\begin{table}[h]
\caption{Cavity eigenmodes of the stage-B5 single-cell cavity with Be windows used for wakefield reconstruction.}
\setlength{\tabcolsep}{6pt} 
\begin{tabular}{lccc}
\toprule
Type & $f_n$ [GHz] & $(R/Q)_n$ [$\Omega$] & $Q_{\mathrm{L},n} [1]$ \\
\hline
TM$_{010}$ & 0.704 & 140.85 & 2.23$\times10^{4}$ \\
TM$_{020}$ & 1.567 & 72.80  &3.23$\times10^{4}$ \\
TM$_{011}$ & 2.169 & 28.70  & 2.45$\times10^{4}$ \\
TM$_{030}$ & 2.373 &  5.44 & 3.52$\times10^4$ \\
TM$_{021}$ & 2.515 & 53.20  & 2.33$\times10^4$ \\
TM$_{022}$ & 3.178 & 17.98  & 2.61$\times10^4$ \\
TM$_{040}$ & 3.226 &  5.25 & 3.90$\times10^4$ \\
TM$_{012}$ & 3.773 & 21.64  & 3.06$\times10^4$ \\
TM$_{031}$ & 3.960 & 1.18  & 3.04$\times10^4$ \\
\toprule
\end{tabular}
\label{tab:eigenmodes}
\end{table}

The results show that, in the short-range wakefield regime, corresponding to the time interval during which the beam is still traversing the cavity, the longitudinal wake potential is dominated by space-charge effects. The DSC component is the predominant term, while the ISC contribution is smaller in magnitude and opposite in sign. Both space-charge components are symmetric with respect to $t=0$, as shown in previous studies~\cite{Schindl1999, MounetISC}. After the bunch has left the cavity, corresponding to the long-range wakefield regime, the space-charge contribution rapidly decreases and becomes negligible. In this region, the wake potential is dominated by the excitation of the cavity eigenmodes, with the fields continuing to resonate within the cavity boundaries. In the long-range regime, the reconstructed eigenmode wake potential shows good agreement with the total wake potential obtained from time-domain simulations, confirming the accuracy of our analytical model used for the reconstruction. While the decomposition of wakefields is illustrated here for the stage-B5 cavity, comparable trends are found across all cavities of each stage. Consequently, the conclusions drawn from this example are representative for all the cavities in the rectilinear cooling channel.

\vspace{0.8em}

\subsection{Transverse dependence of wake potential in RF cavities with and without windows}
\label{transv_dep_R_Q}
\begin{figure}[hbt!]
    \includegraphics[width=0.48\textwidth]{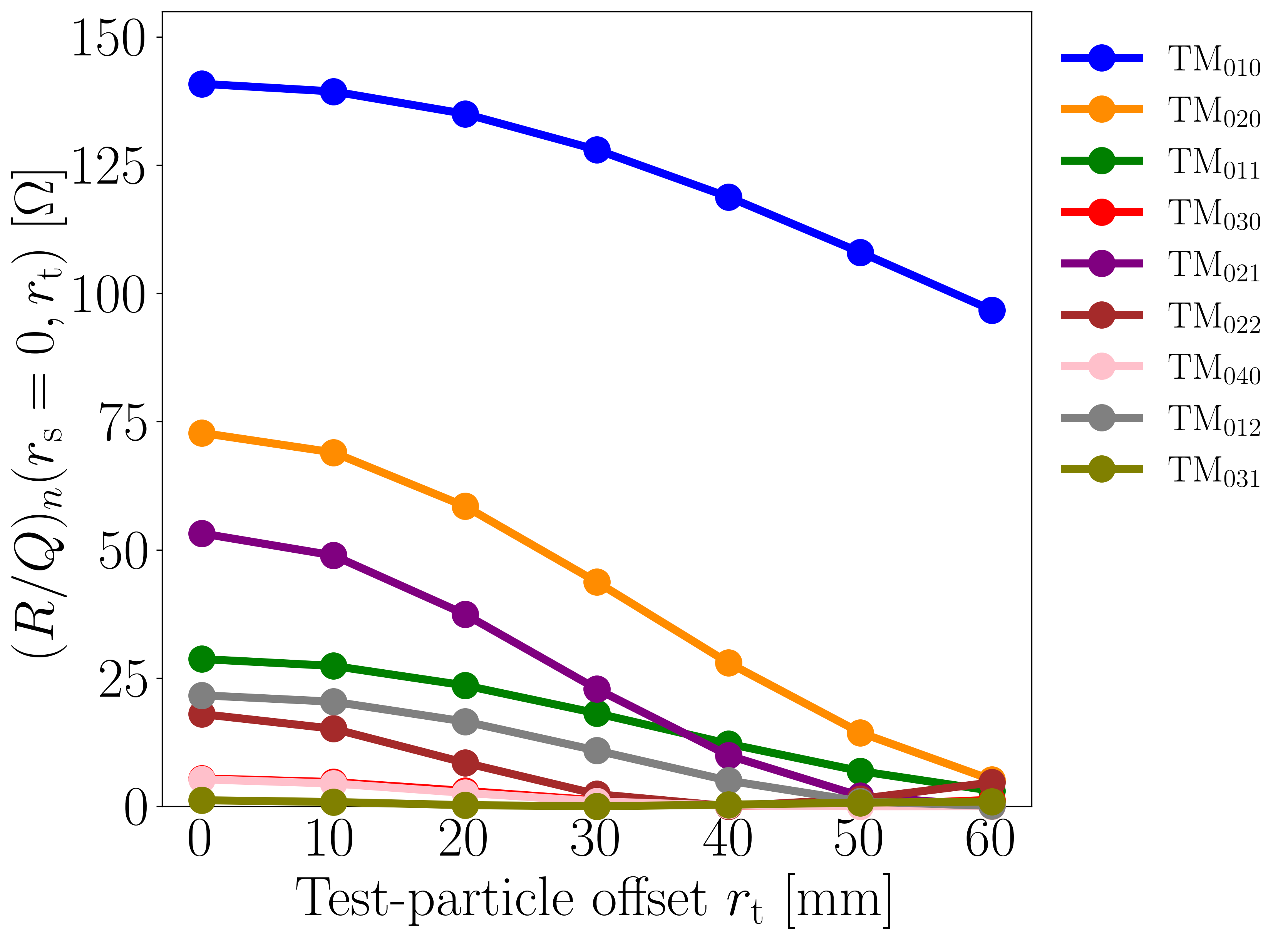}
    \caption{Variation of the generalized $(R/Q)_{n}$ of stage-B5 single-cell cavity as a function of test-particle transverse offset $r_\mathrm{t}$ for the relevant monopole modes.}
    \label{fig:RQ_rt_modes}
\end{figure}

In contrast to the conventional cavities, where the beam pipe is entirely in vacuum, in cavities equipped with Be windows, the longitudinal beam-cavity interaction depends on the transverse offset of the source and test particles, even for monopole modes. To account for this effect, the on-axis geometric shunt impedance can be replaced by the generalized $(R/Q)_n$ evaluated at the radial positions of the source ($r_\mathrm{s}$) and test ($r_\mathrm{t}$) particles:

\begin{equation}
\begin{aligned}
(R/Q)_{n}(r_\mathrm{s},r_\mathrm{t}) &= \frac{\left| V_{z,n}(r_\mathrm{s}) V_{z,n}^{*}(r_\mathrm{t}) \right|}{\omega_{n} U_{n}} \\
                    &= \sqrt{(R/Q)_{n}(r_\mathrm{s})} \; \sqrt{(R/Q)_{n}(r_\mathrm{t})} \,,
\end{aligned}
\label{eq:RQ_rt_rs}
\end{equation}
\noindent
where $V_{z,n}(r_\mathrm{s})$ is the longitudinal voltage of mode $n$ at the position $r_\mathrm{s}$, $V_{z,n}^{*}(r_\mathrm{t})$ the complex conjugate of the longitudinal voltage of mode $n$ at the position $r_\mathrm{t}$, and $\omega_n$ and $U_n$ are the angular frequency and stored energy of the considered mode, respectively. The modal factorization in Eq.~\eqref{eq:RQ_rt_rs} is valid only for axisymmetric TM monopole modes, which are relevant for our study. Since our cavity geometry is perfectly rotationally symmetric and enclosed by Be windows in the irises without port openings, higher-order azimuthal multipoles ($m \ge 1$) are absent, and the longitudinal beam-cavity interaction is strictly governed by pure $m=0$ modes. In the specific case, the source particle is placed on axis ($r_\mathrm{s}=0$), and the test particle varies from $r_\mathrm{t}=0$ to the radius of the cavity window $R_\mathrm{i}$. The dependence of the generalized $(R/Q)_{n}$ of stage-B5 single-cell cavity on the radial position of the test particle  $r_\mathrm{t}$ is shown in Fig.~\ref{fig:RQ_rt_modes}.

After establishing the separation of the longitudinal wake potential into its FM, HOMs, and SC components for an RF cavity closed by windows (see Sec.~\ref{wakefield_section}), in the following section, we describe the behavior of these components in multi-cell cavities with and without windows, taking the Stage-B5 as a representative example. For the windowless multi-cell cavity, the reconstructed wake potential accounts for the trapped TM monopole modes below the $\mathrm{TM_{01}}$ beam-pipe cutoff frequency ($f_{c,\mathrm{TM}_{01}} \approx 1.91\mathrm{~GHz}$ for $R_{\mathrm{i}} = 60\mathrm{~mm}$), corresponding to the first two monopole passbands up to approximately $1.69\mathrm{~GHz}$.

\begin{figure*}[hbt!]
\centering

\begin{subfigure}{0.325\textwidth}
    \includegraphics[width=\linewidth]{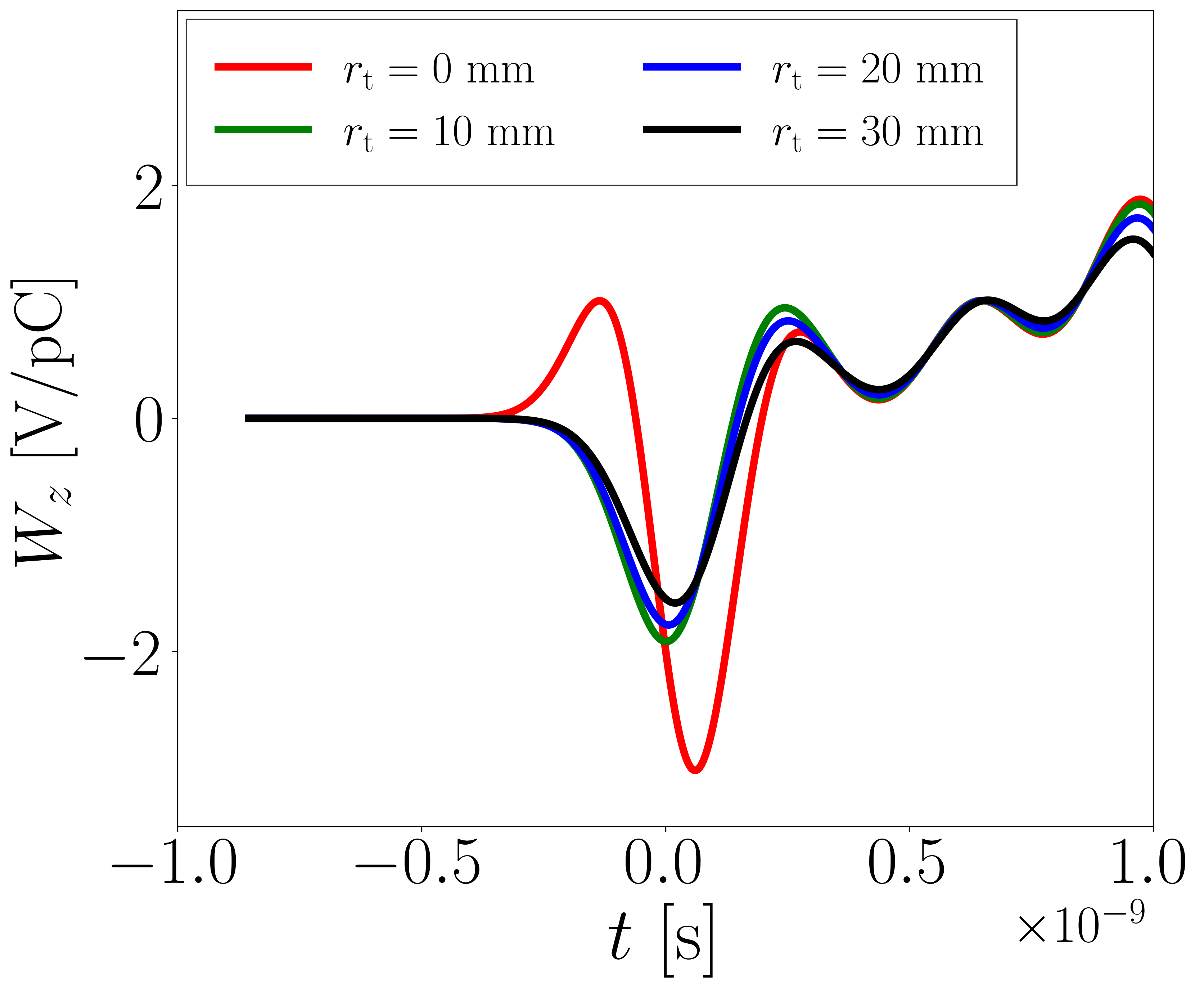}
    \caption{$W_{z,\mathrm{tot}}(t)$}
    \label{fig:Wz_tot_Be_windows}
\end{subfigure}
\hfill
\begin{subfigure}{0.325\textwidth}
    \includegraphics[width=\linewidth]{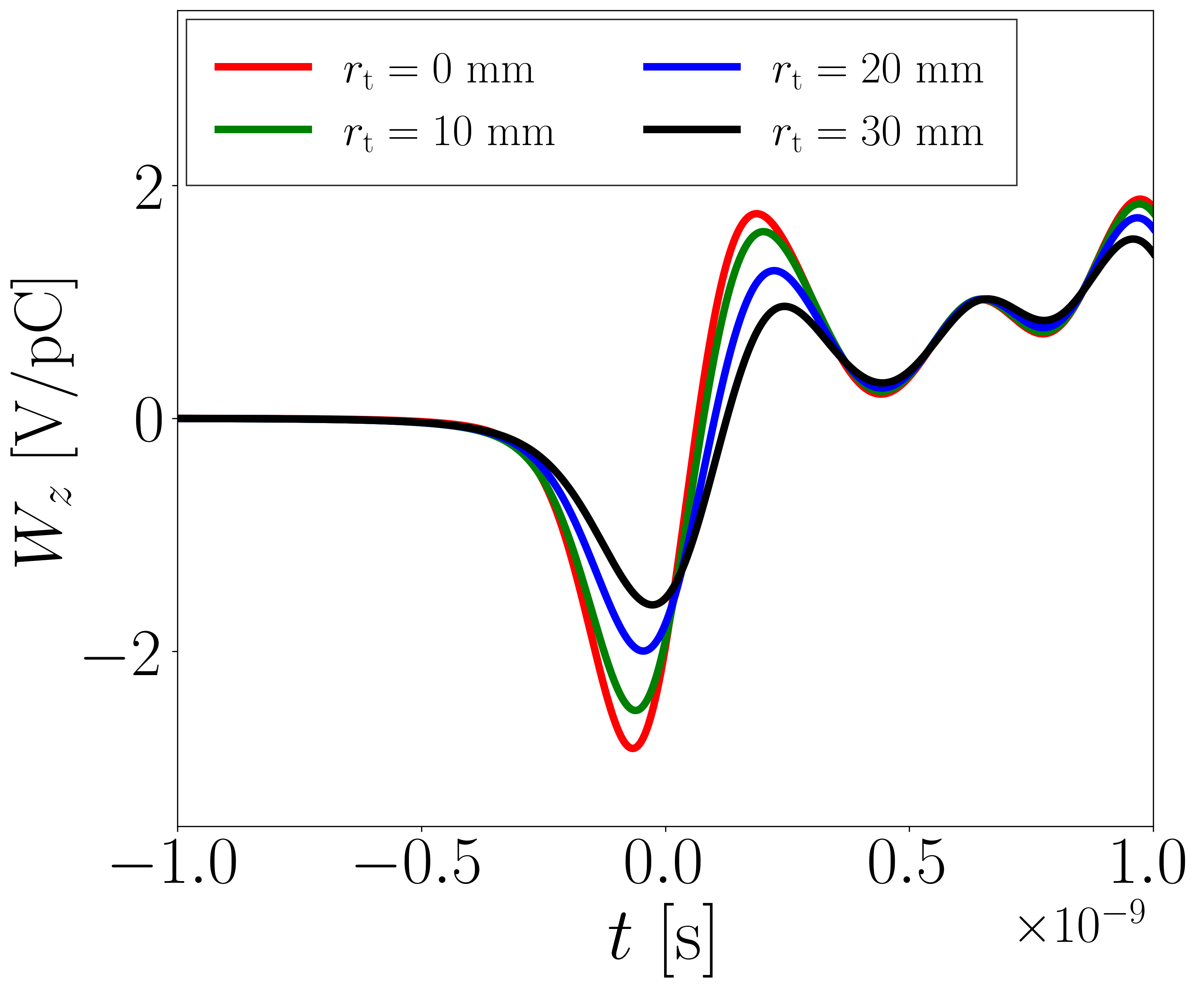}
    \caption{$W_{z,\mathrm{FM}}(t)+W_{z,\mathrm{HOMs}}(t)+W_{z,\mathrm{ISC}}(t)$}
     \label{fig:Wz_FM_HOMs_ISC_Be_windows}
\end{subfigure}
\hfill
\begin{subfigure}{0.325\textwidth}
    \includegraphics[width=\linewidth]{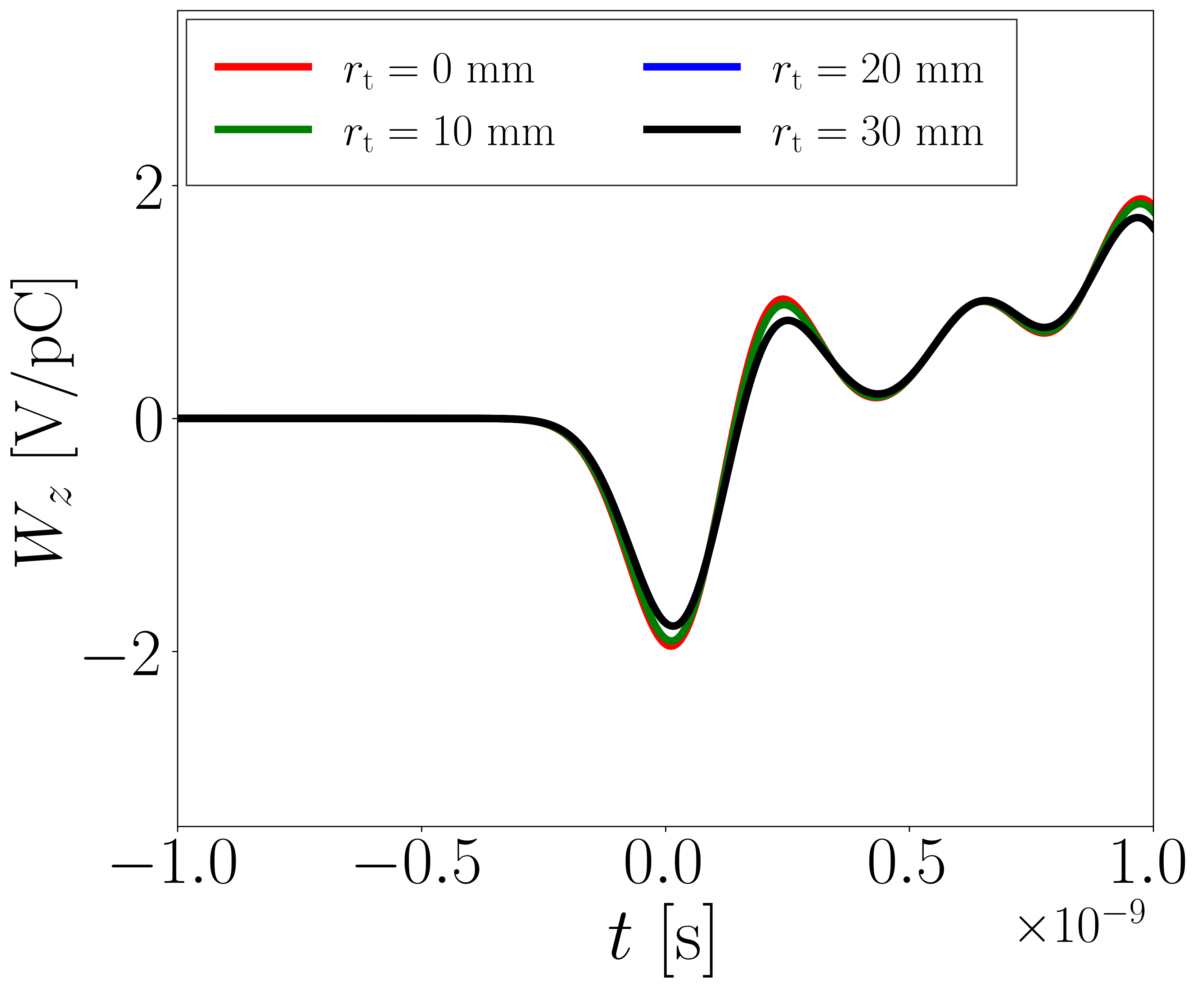}
    \caption{$W_{z,\mathrm{FM}}(t)+W_{z,\mathrm{HOMs}}(t)$}
    \label{fig:Wz_FM_HOMs_Be_windows}
\end{subfigure}

\vspace{0.4cm}

\begin{subfigure}{0.325\textwidth}
    \includegraphics[width=\linewidth]{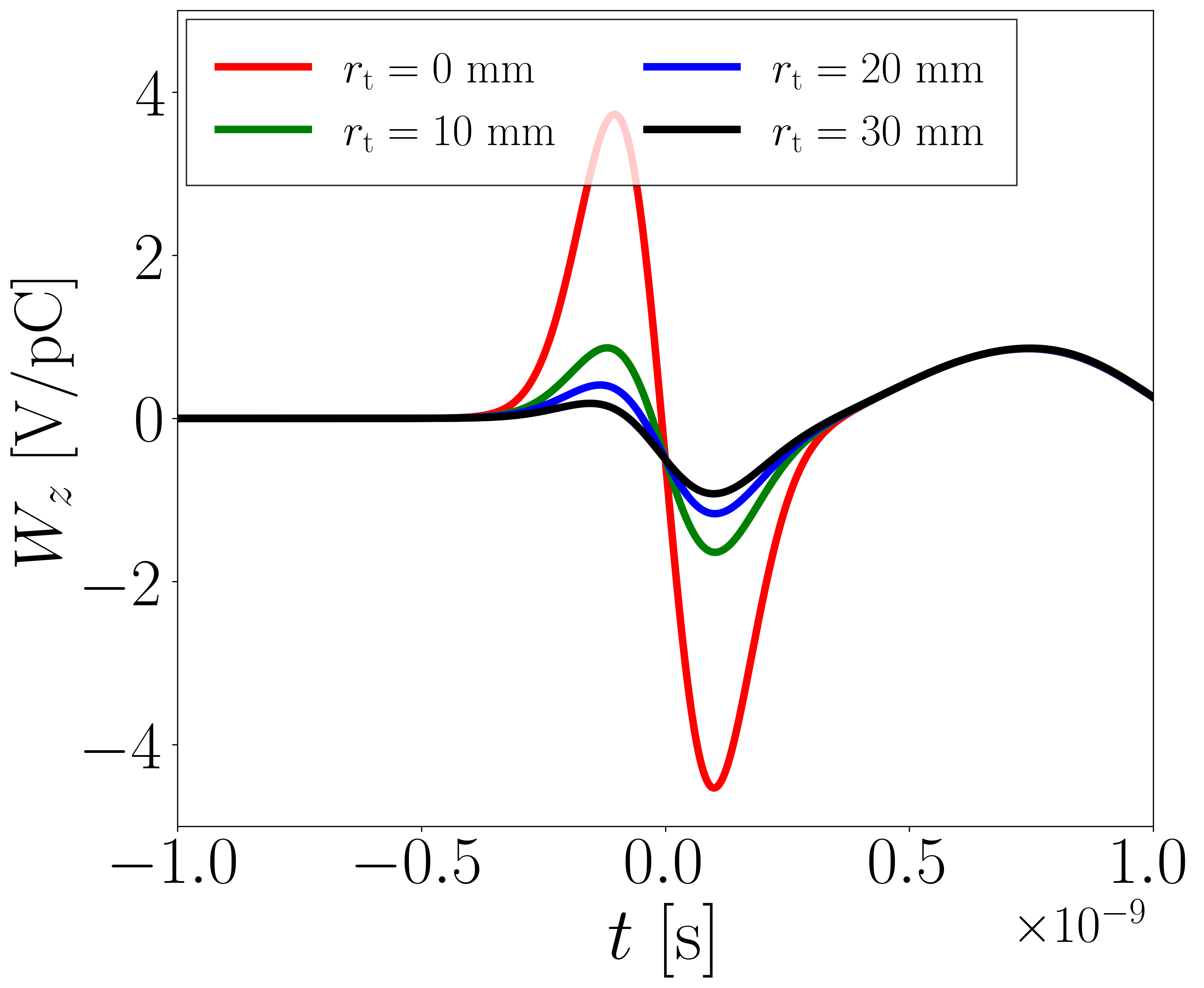}
    \caption{$W_{z,\mathrm{tot}}(t)$}
    \label{fig:Wz_tot_no_Be_windows}
\end{subfigure}
\hfill
\begin{subfigure}{0.325\textwidth}
    \includegraphics[width=\linewidth]{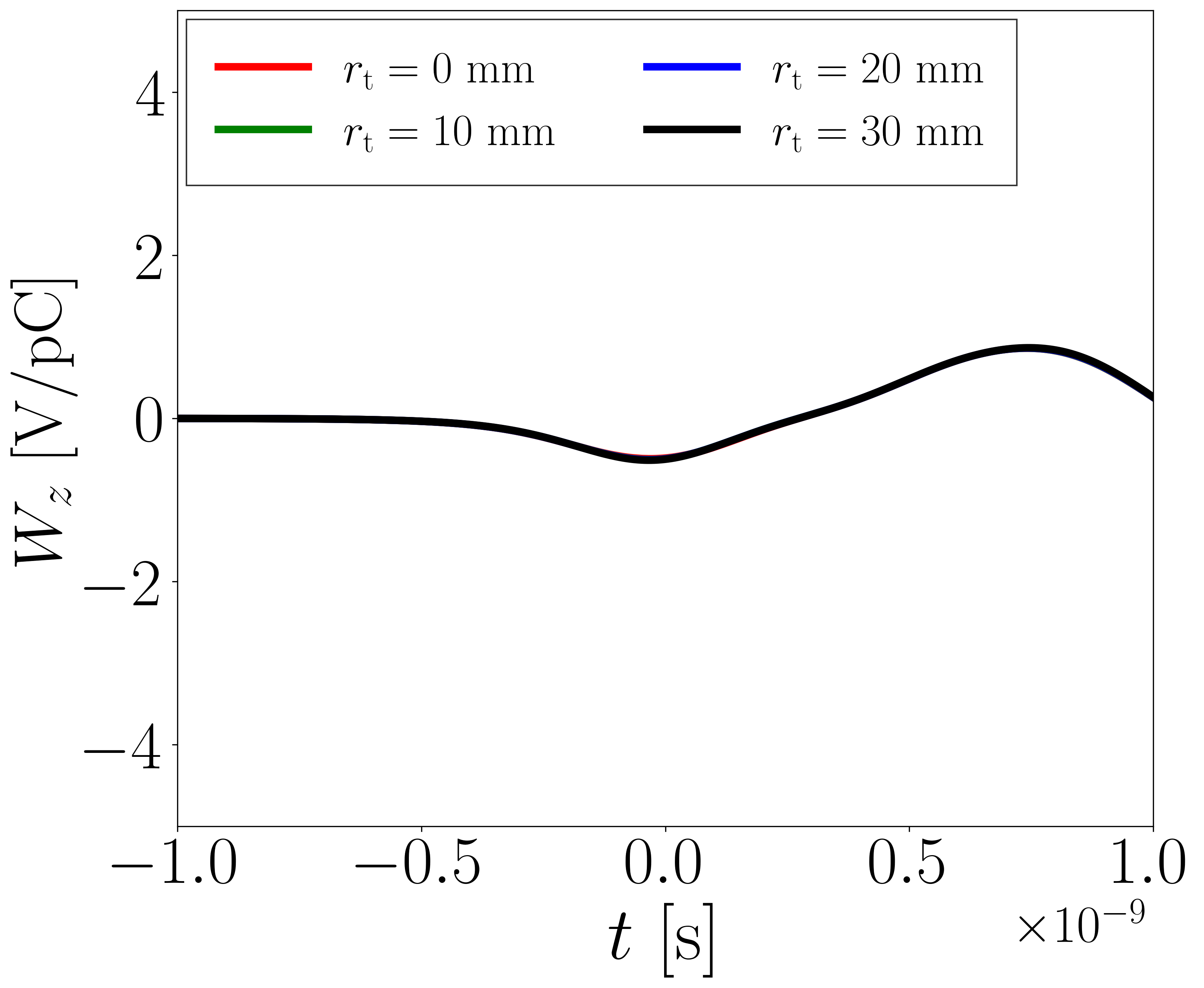}
    \caption{$W_{z,\mathrm{FM}}(t)+W_{z,\mathrm{HOMs}}(t)+W_{z,\mathrm{ISC}}(t)$}
    \label{fig:Wz_FM_HOMs_ISC_No_Be_windows}
\end{subfigure}
\hfill
\begin{subfigure}{0.325\textwidth}
    \includegraphics[width=\linewidth]{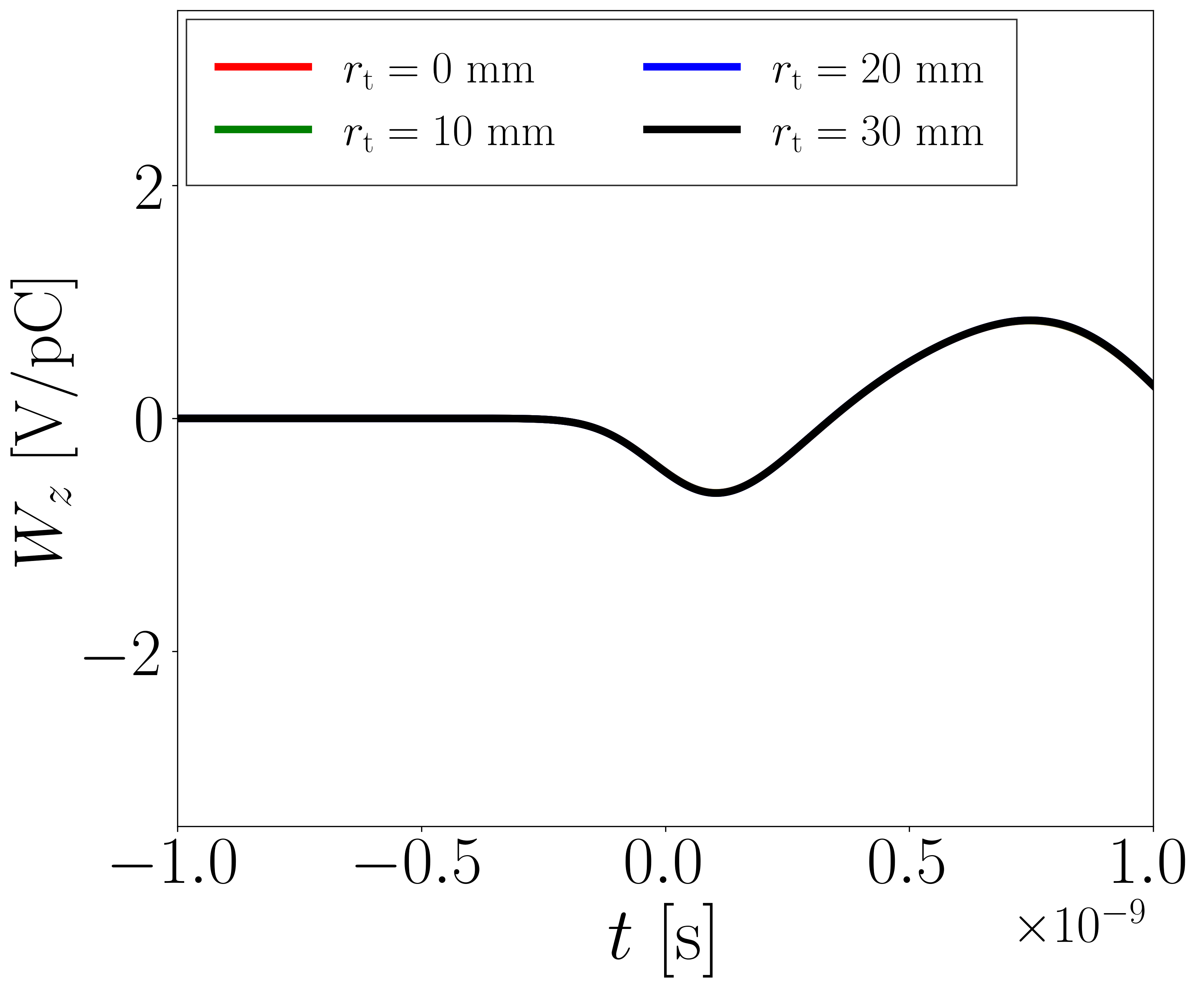}
    \caption{$W_{z,\mathrm{FM}}(t)+W_{z,\mathrm{HOMs}}(t)$}
      \label{fig:Wz_FM_HOMs_No_Be_windows}
\end{subfigure}
\caption{Decomposition of the longitudinal wake potential as a function of the test-particle transverse offset $r_\mathrm{t}$ for the stage-B5 multi-cell cavity with Be windows (top) and without Be windows (bottom).
}
\label{fig:wake_decomposition}
\end{figure*}

\begin{figure}[hbt!]
    \centering

    \begin{subfigure}{0.4\textwidth}
        \centering
        \includegraphics[width=\textwidth]{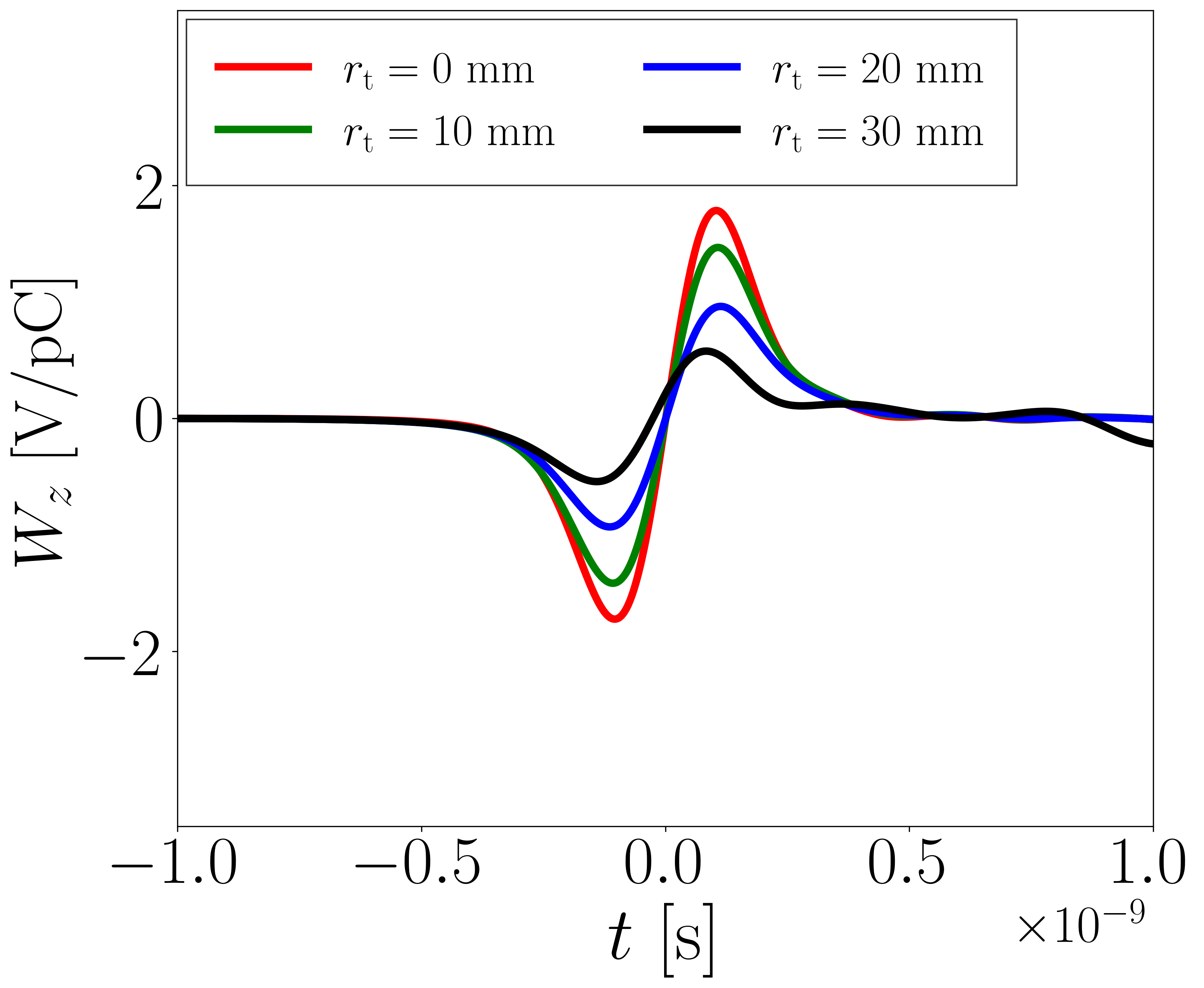}
        \caption{$W_{z,\mathrm{ISC}}(t)$}
        \label{fig:Wz_ISC_wind}
    \end{subfigure}

    \vspace{0.5cm}

    \begin{subfigure}{0.4\textwidth}
        \centering
        \includegraphics[width=\textwidth]{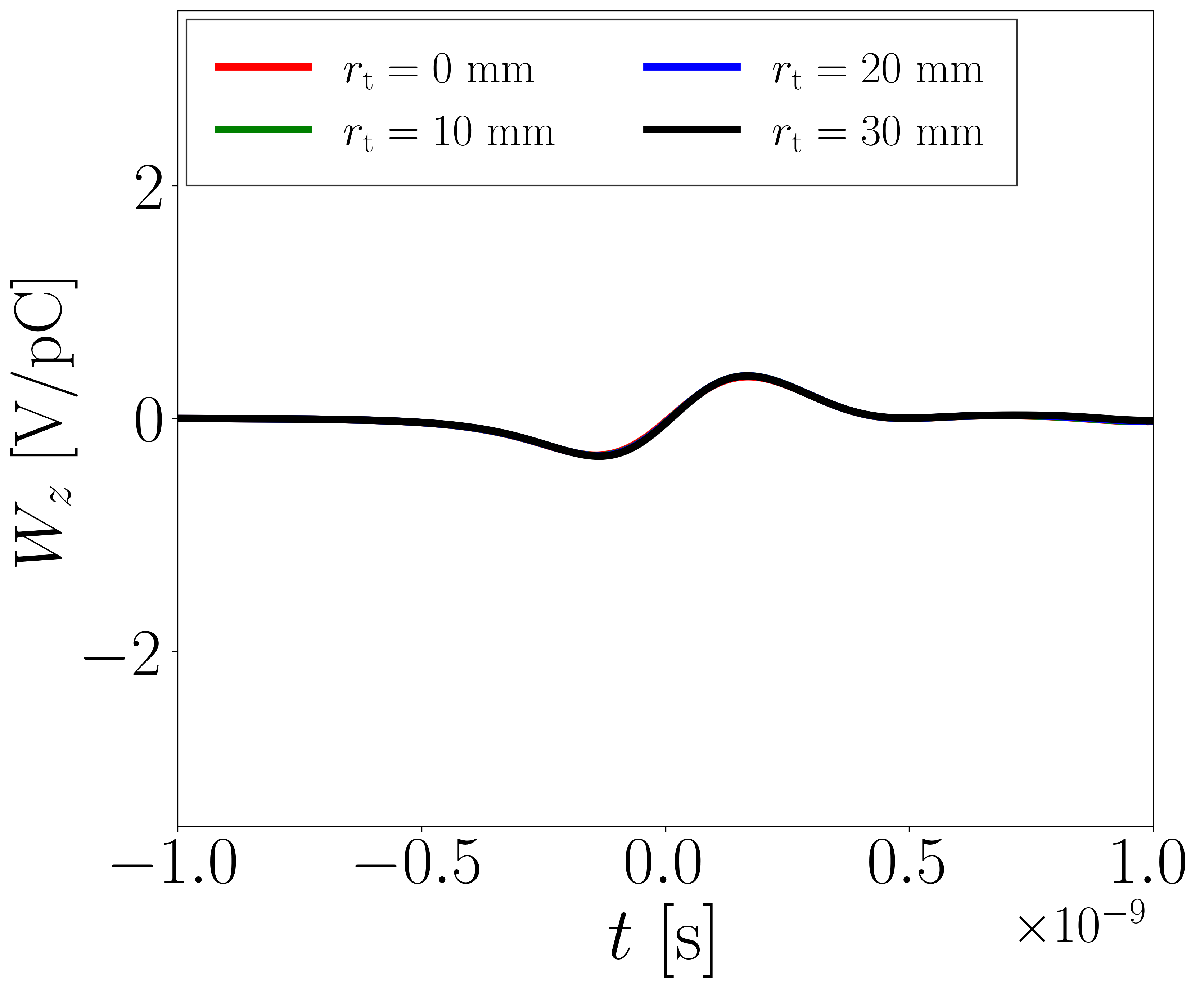}
        \caption{$W_{z,\mathrm{ISC}}(t)$}
        \label{fig:Wz_ISC_no_wind}
    \end{subfigure}

    \caption{Indirect space-charge longitudinal wake potential as a function of the test-particle transverse offset $r_\mathrm{t}$ for the stage-B5 multi-cell cavity with Be windows (a) and without Be windows (b).}
    \label{fig:Wake_ISC}
\end{figure}

Figure~\ref{fig:wake_decomposition} shows, from left to right, the total wake potential $W_{z,\mathrm{tot}}(t)$, the cavity-dependent contribution $W_{z,\mathrm{FM}}(t) + W_{z,\mathrm{HOMs}}(t)+W_{z,\mathrm{ISC}}(t)$, and the reconstructed eigenmode wake potential $W_{z,\mathrm{FM}}(t) + W_{z,\mathrm{HOMs}}(t)$ as a function of the test-particle transverse offset $r_{\mathrm{t}}$ for the multi-cell stage-B5 cavity with Be windows (top) and without windows (bottom). The isolated ISC wake potential contributions for the cavity configurations with and without windows are plotted in Fig.~\ref{fig:Wz_ISC_wind} and Fig.~\ref{fig:Wz_ISC_no_wind}, respectively. The DSC contributions are not shown here, as their transverse dependence was already discussed in Section~\ref{wakefield_section} and illustrated in Fig.~\ref{fig:Wake_BP}.

For the cavity with Be windows, the total wake potential $W_{z,\mathrm{tot}}(t)$ (see Fig.~\ref{fig:Wz_tot_Be_windows}) exhibits a nonlinear dependence on the transverse offset of the test particle, especially in the short-range wakefield regime. In addition, a residual offset effect is also observed in the long-range regime. This effect is due to the radial dependence of the longitudinal electric field of FM and HOMs, and consequently of their longitudinal $(R/Q)_{n}$ values, induced by the presence of the Be windows. 

For the cavity without windows, the total wake potential (see Fig.~\ref{fig:Wz_tot_no_Be_windows}) shows a nonlinear sensitivity to $r_{\mathrm{t}}$ only in the short-range wake region. Conversely, the long-range wake potential is independent of the transverse offset since the modes can freely ring inside the cavity in the absence of obstacles, i.e., the windows, along the beam axis. It is also worth noting that the reduced short-range wake amplitude in the cavity with Be windows indicates that windows act as effective dampers of total space-charge effects.  

The cavity-dependent wake potential, $W_{z,\mathrm{FM}}(t) + W_{z,\mathrm{HOMs}}(t)+W_{z,\mathrm{ISC}}(t)$, obtained by subtracting the DSC wake potential described in Eq.~\eqref{eq:WDSC_cavity} from the total wake potential, exhibits a clear offset dependence in both wakefield regimes for the cavity with Be windows, as shown in Fig.~\ref{fig:Wz_FM_HOMs_ISC_Be_windows}. In the short-range regime, this effect is mainly associated with the ISC contribution, since the DSC term has been removed. In the long-range regime, where space-charge effects are negligible, the transverse dependence is dominated by the $(R/Q)_{n}$ values' sensitivity on the offset $r_{\mathrm{t}}$. Conversely, in the absence of Be windows, the cavity-dependent wake potential does not exhibit any transverse sensitivity and is governed mainly by the longitudinal mode excitation (see Fig.~\ref{fig:Wz_FM_HOMs_ISC_No_Be_windows}). 

The eigenmode wake potential $W_{z,\mathrm{FM}}(t) + W_{z,\mathrm{HOMs}}(t)$, reconstructed from the modal expansion, further clarifies this behavior. It shows a nonlinear dependence on $r_{\mathrm{t}}$ in the presence of windows (see Fig.~\ref{fig:Wz_FM_HOMs_Be_windows}), reflecting the radial dependence of FM and HOMs electric fields and their $(R/Q)_{n}$ values. No such sensitivity is observed for the cavity without windows (see Fig.~\ref{fig:Wz_FM_HOMs_No_Be_windows}). The ISC contribution $W_{z,\mathrm{ISC}}(t)$, obtained by subtracting the eigenmode wake potential from the cavity-dependent wake potential, shows a transverse sensitivity only in the short-range wakefield regime (see Fig.~\ref{fig:Wz_ISC_wind}) in the cavity with windows. No such dependence is detected in the absence of windows (see Fig.~\ref{fig:Wz_ISC_no_wind}), in agreement with the behavior expected for a conventional cavity with a beam pipe.

Overall, these results demonstrate that the presence of the windows introduces a radial dependence in the wake contributions associated with FM and HOM excitation and the ISC. In contrast, in the absence of windows, both contributions are essentially independent of the transverse offset, and the observed sensitivity of the total wake potential arises mainly from the DSC term. An important implication is that, in cavities with windows, both the DSC and ISC wake potentials depend nonlinearly on the transverse offset, so a realistic transverse distribution of the beam must be taken into account. Consequently, the wakefield solver results do not accurately capture the space-charge effects for a realistic beam distribution. Therefore, these effects are excluded from consideration for beam-loading compensation. Other tools, such as PIC codes, must be used to correctly assess the space-charge effects. Furthermore, the FM and HOM effects can be mitigated by beam-loading compensation techniques only at a single transverse location, for example, on axis. The details and results of the beam-loading compensation strategy for the studied cavities are discussed in Section~\ref{beam_loading}.

\subsection{Eigenmode wake potential for finite transverse beam size}
\label{twdwp}

Equations~\eqref{eq:Ws_total} and \eqref{eq:convolution} define, respectively, the pencil-beam longitudinal wake function and wake potential reconstructed from the eigenmode solution. In this case, the resulting wake potential is equivalent to the one generated by monopole modes when excited by a longitudinal bunch charge distribution with zero transverse size. Consequently, only the on-axis field of each mode contributes to the beam–cavity interaction, which is characterized by their respective on-axis geometric shunt impedance, $(R/Q)_n$. However, a realistic muon bunch has a finite transverse size and interacts with the radial field profile of each cavity mode. This effect results from the presence of Be windows on the irises of the muon cooling cavities, as discussed in Section~\ref{transv_dep_R_Q}. In the following, we describe a method to include finite transverse-beam size effects in the reconstruction of the longitudinal eigenmode wake potential.

For simplicity, we assume a radially symmetric transverse bunch distribution, i.e., the rms beam size $\sigma_{\mathrm{t}}$ is the same in both transverse directions. In this case, the differential charge contained in an annular ring at radius $r_\mathrm{s}$ is defined as

\begin{equation}
dQ_b(r_\mathrm{s})=Q_b\,2\pi r_\mathrm{s}\,\rho_r(r_\mathrm{s})\,dr_\mathrm{s},
\end{equation}
\noindent
where $\rho_r(r_\mathrm{s})$ is the radial Gaussian charge density, truncated at the window radius $R_{\mathrm{i}}$:

\begin{equation}
\rho_r(r_\mathrm{s})
=
\frac{1
}{
2\pi\displaystyle\int_{0}^{R_{\mathrm{i}}}
r'\exp\!\left(-\frac{{r'}^{2}}{2\sigma_\mathrm{t}^{2}}\right)dr'
}\exp\!\left(-\frac{r_\mathrm{s}^{2}}{2\sigma_\mathrm{t}^{2}}\right)
,
\end{equation}

\noindent
with the normalization condition

\begin{equation}
\int_{0}^{R_{\mathrm{i}}} 2\pi r_\mathrm{s}\,\rho_r(r_\mathrm{s})\,dr_\mathrm{s} = 1.
\end{equation}

\begin{figure}[hbt!]
    \centering

    \begin{subfigure}{0.45\textwidth}
        \centering
        \includegraphics[width=\textwidth]{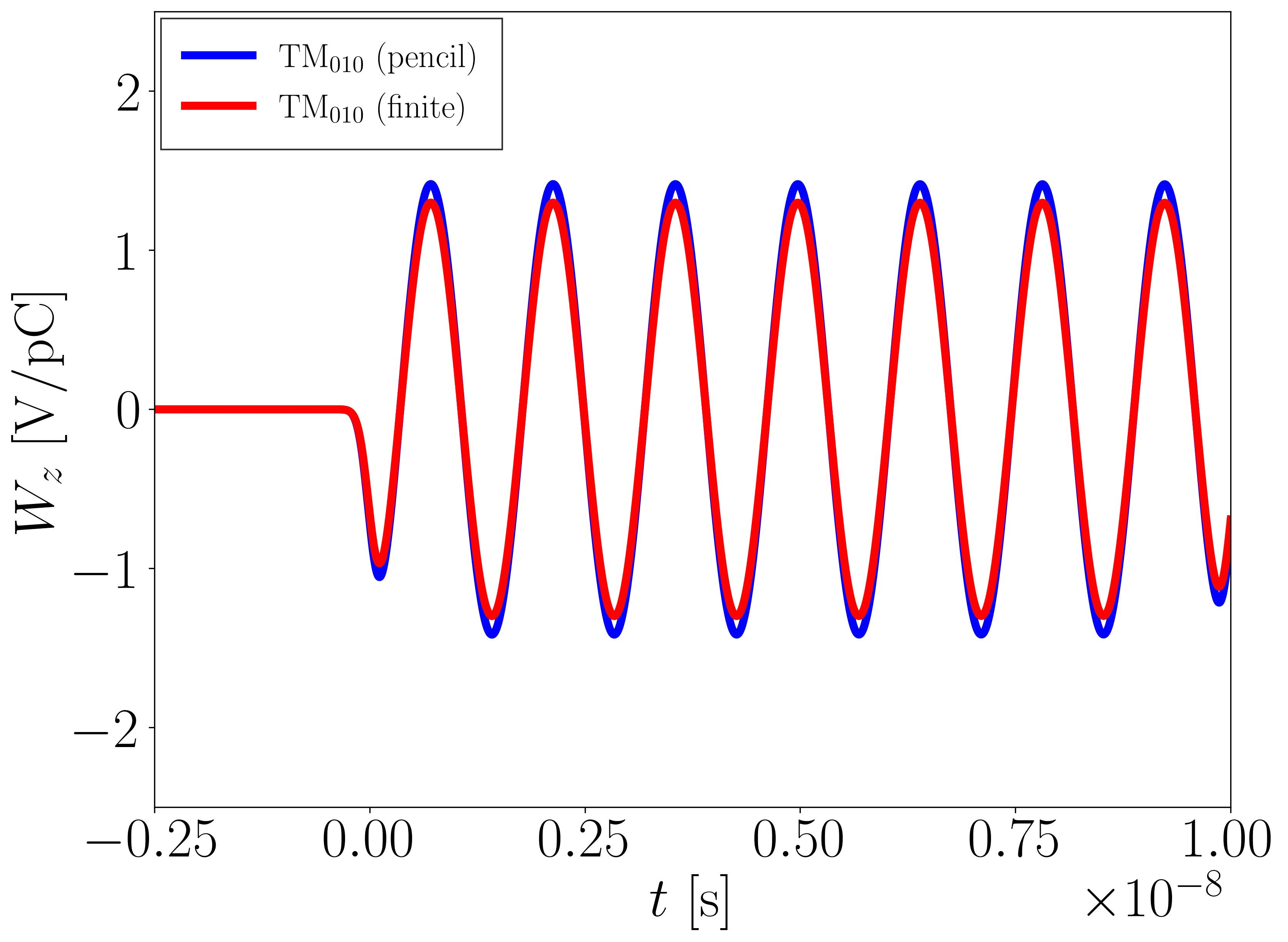}
        \caption{}
        \label{fig:Wz_FM}
    \end{subfigure}

    \vspace{0.5cm}  

    \begin{subfigure}{0.45\textwidth}
        \centering
        \includegraphics[width=\textwidth]{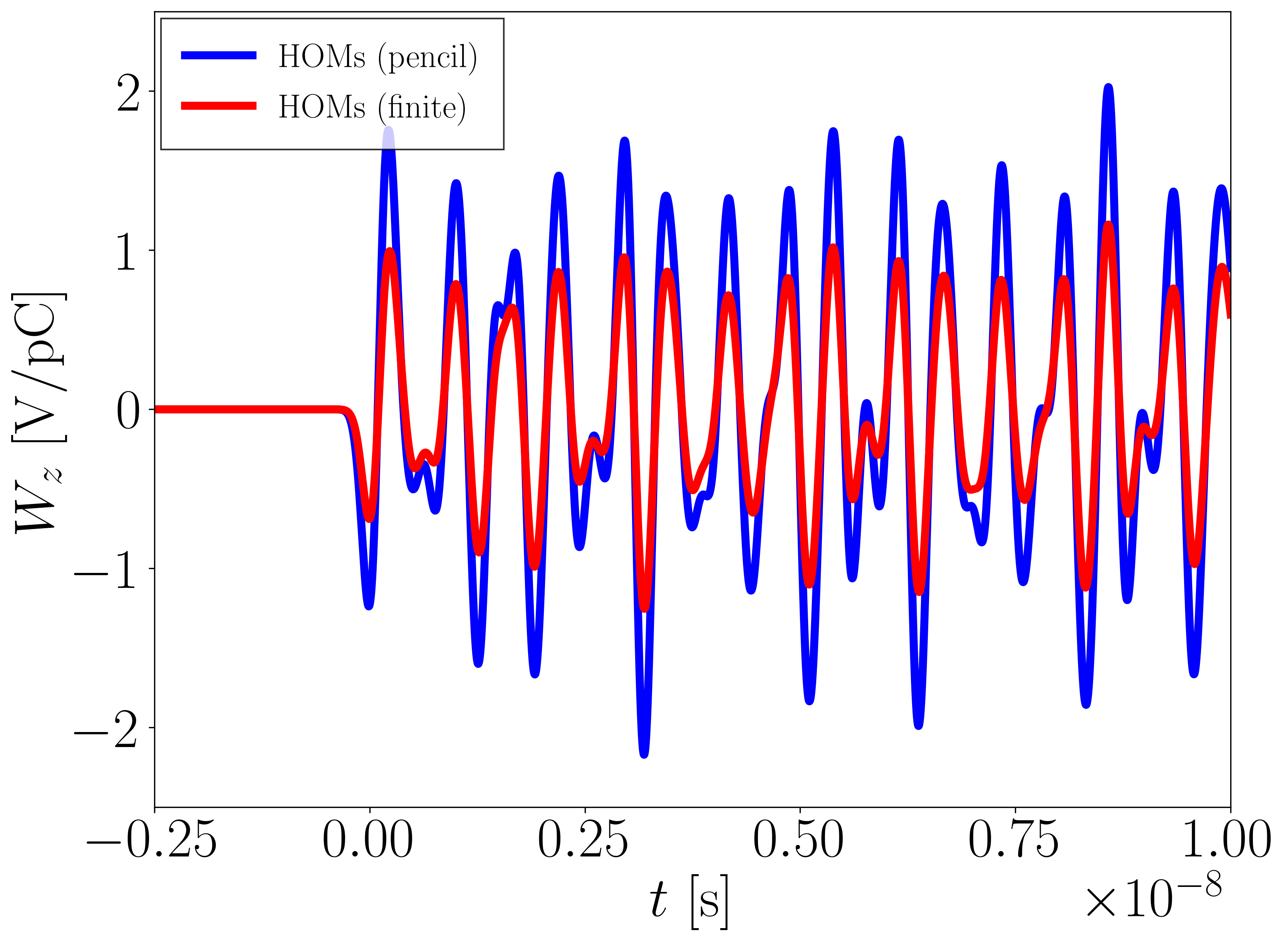}
        \caption{}

        \label{fig:Wz_HOMs}
    \end{subfigure}

    \caption{Longitudinal wake potentials of the FM (a) and the HOMs listed in Table~\ref{tab:eigenmodes} (b) excited by a pencil beam (blue) and a finite beam with a transverse size of $\sigma_\mathrm{t}$ = 20~mm (red) in the stage-B5 multi-cell cavity with Be windows.}
    \label{fig:{Pencil_vs_finite_wake_potential}}
\end{figure}

Using Eq.~\eqref{eq:RQ_rt_rs} and integrating the differential contributions from all annular rings, we can define a mode-dependent transverse form factor as
\begin{equation}
F_{n,\perp}(\sigma_{\mathrm{t}})
=\int_{0}^{R_{\mathrm{i}}} 2\pi r_\mathrm{s}\,\rho_r(r_\mathrm{s})\,
\sqrt{(R/Q)_{n}(r_\mathrm{s})}\,dr_\mathrm{s}.
\label{eq:F_perp}
\end{equation}
Multiplying the wake function (Eq.~\eqref{eq:Ws_total}) by this transverse form factor and performing the convolution with the longitudinal bunch charge distribution \(\lambda(t)\) yields the finite-beam eigenmode wake potential 

\begin{equation}
W_{z,\mathrm{eig}}^{\mathrm{finite}}(t, \sigma_\mathrm{t})
= \frac{1}{Q_b} \int_{-\infty}^{t} F_{n,\perp}(\sigma_{\mathrm{t}}) W_{z,\mathrm{eig}}^{0}(t-t') \lambda(t') \, dt',
\label{eq:W_eig_gauss_transverse}
\end{equation}
\noindent
where $\lambda(t)$ is the longitudinal Gaussian density of Eq.~\eqref{eq:lambda_gaussian}. This expression accounts for the wake excitation produced by a bunch with non-zero transverse size. The resulting FM and HOMs wake potentials, induced by both a pencil beam and a finite-size beam, with $\sigma_\mathrm{t}$~=~20~mm, in the stage-B5 multi-cell cavity with Be windows, are shown in Fig.~\ref{fig:Wz_FM} and Fig.~\ref{fig:Wz_HOMs}, respectively.

The fundamental mode ($\mathrm{TM_{010}}$) is weakly affected by the inclusion of the transverse charge distribution as its longitudinal electric field is nearly uniform across the cavity aperture. Conversely, the HOM wake exhibits higher suppression due to the stronger radial dependence of the HOM fields and their corresponding $(R/Q)_n$ (see Fig.~\ref{fig:RQ_rt_modes}). Overall, accounting for the transverse beam size reduces the wake potential amplitude, with the magnitude of the effect determined by the radial distribution of the eigenmode fields. 

\subsection{Benchmark of the analytical wakefield reconstruction with PIC simulations}

The methodology described in the previous sections models the longitudinal wakefield contributions when excited by a charged particle bunch with a finite transverse size. Standard wakefield solvers, such as the one available in CST Studio Suite, are limited to wakefield excitations by a rigid line charge and therefore cannot directly account for transverse beam distributions. To overcome this limitation, the PIC solver of CST can be used to model the full three-dimensional bunch distribution and to directly solve for the electromagnetic fields generated by the bunch traversing the cavity. 
To validate the analytical wakefield-based approach that accounts for the finite transverse beam size discussed in Sec.~\ref{twdwp}, a dedicated post-processing workflow was implemented in Python to reconstruct the longitudinal wakefields from the electric field data exported by the PIC solver of CST for the stage-B5 single-cell cavity.

The generalized longitudinal wake potential for an on-axis probe particle at a longitudinal distance $s = \beta c t - z$ from the bunch center is given by
\begin{equation}
    W_z (s) = \frac{\beta c}{Q_b} \int_{-\infty}^{\infty}E_{z,\text{PIC}}(\beta c t - s,t)\big\rvert_{\beta c t = s + z }dt,
    \label{wake_z_pic}
\end{equation}
where $Q_b$ is the total bunch charge and $E_{z,\text{PIC}}$ is the longitudinal electric field obtained from the PIC simulation, evaluated along a prescribed reference trajectory. Here, $s$ denotes the longitudinal coordinate in the co-moving frame, while $t$ and $z$ are the time and longitudinal position in the laboratory frame.

\begin{figure}[htbp]
    \centering
    \includegraphics[width=0.65\linewidth]{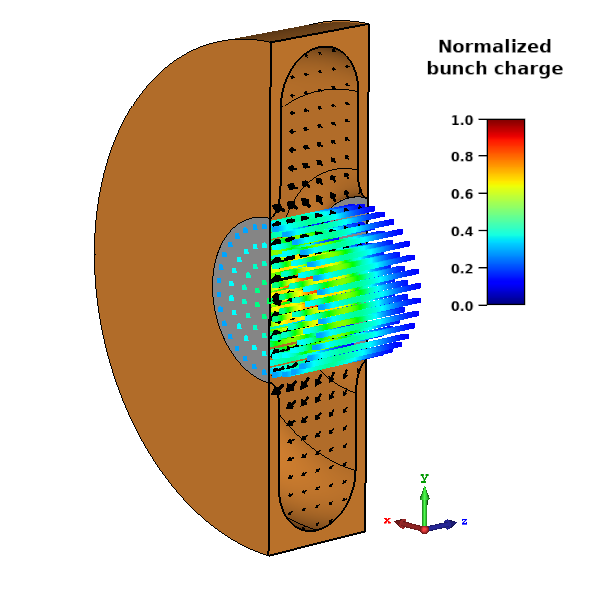}
    \caption{Example setup for a PIC solver simulation. The particle source is released from the exterior wall of the Be window and traverses the cavity along its axis. The particle bunch excites electromagnetic fields within the cavity, with the black arrows representing the electric field vector.}
    \label{fig:PIC-B5-geom}
\end{figure}

Figure \ref{fig:PIC-B5-geom} illustrates a representative PIC configuration for the stage-B5 single-cell cavity with Be windows. The particle source is located on the outer window wall, which serves as the cathode for particle injection. The thin Be window is treated as transparent to particles, meaning that in the PIC model particles traverse the material without undergoing atomic interactions such as scattering or energy loss.

In the simulation, the particle source utilizes the stage-B5 beam parameters from Table~\ref{tab:beam_dynamics_parameters}, with the bunch transverse dimension set to $R_{\mathrm{i}}/3$. The charge of the particle bunch is initialized with Gaussian spatial distributions in both the longitudinal and transverse directions. All particles are assigned the same purely longitudinal momentum, with the transverse momentum components set to zero. Due to its relatively large mass, the muon bunch satisfies the rigid-beam approximation required for the accurate computation of the wakefields.

All six boundaries of the simulation domain are specified as PEC. The computational mesh comprises approximately 24,000 cells, with the minimum cell size of 770~\textmu m, chosen to be smaller than the characteristic distance traveled by a particle within a single time step, 2.7~mm. The total simulation duration is required to be at least as long as the time needed for the final test particle, in the co-moving reference frame, to traverse the entire integration region used for post-processing the wakefield. The post-processing procedure treats the co-moving coordinate as an independent variable. For each time step of the simulation, the test particle’s position in the laboratory frame is evaluated along the cavity axis. The corresponding value of the electric field at those coordinates is then extracted from the CST simulation data. The sequence of longitudinal electric fields sampled in this manner is numerically integrated using Gaussian quadrature~\cite{GQ} to compute the wakefield as defined by Eq.~\eqref{wake_z_pic}. 

\begin{figure}[htb!]
    \centering
    \includegraphics[width=\linewidth]{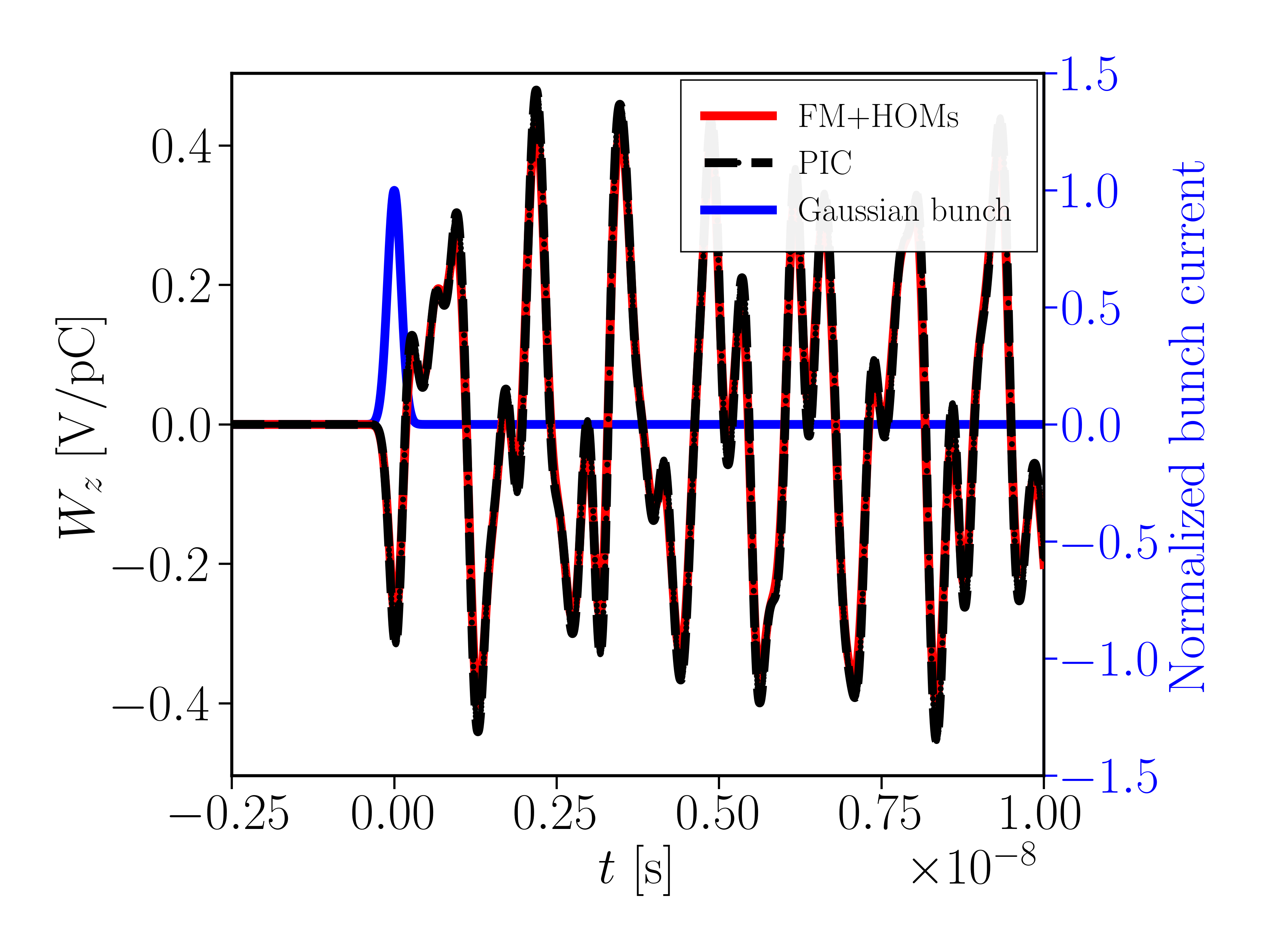}
    \caption{Comparison of the on-axis longitudinal wake potential excited by a bunch with finite transverse size using the analytical eigenmode method and the PIC solver for the stage-B5 single-cell cavity with Be windows. The normalized bunch distribution is shown on the secondary axis.}
    \label{fig:PIC-wake}
\end{figure}

Figure \ref{fig:PIC-wake} compares the on-axis longitudinal wake potential obtained from the analytical eigenmode model described by Eq.~\eqref{eq:W_eig_gauss_transverse} with that computed using the PIC solver for the stage-B5 single-cell cavity with Be windows. The resulting PIC wake potential formally represents a superposition of the wakes generated by the charges present in the cavity (i.e., the drive bunch and its associated image charges) and the wakes induced by the excited cavity modes.

The excellent agreement of the two curves validates the analytical eigenmode decomposition method developed to account for a finite transverse beam size.

\section{Beam Loading and Compensation}\label{sec:4}
\label{beam_loading}
After describing the decomposition of the longitudinal wake potential into its resonant and space-charge components, we now use only the resonant contribution to describe the corresponding beam-induced voltage and discuss its impact on beam loading in the cavities with Be windows of the rectilinear muon cooling channel. Beam loading occurs when the voltage induced by the beam superimposes on the externally applied RF voltage, thereby modifying both its amplitude and phase~\cite{wangler2008rf}.

The total voltage experienced by a beam traversing a cavity is the sum of the externally applied RF voltage and the beam-induced voltage: 

\begin{equation}
V(t) = \underbrace{V_{\mathrm{RF,0}}\sin(\omega_{\mathrm{0}}t+\phi_{\mathrm{0}})}_{V_{\mathrm{RF}}(t)} + V_{b}(t),
\label{eq:V_seen_beam}
\end{equation}

\noindent
where $V_{\mathrm{RF,0}}=TTFE_\mathrm{\mathrm{acc}}L_{\mathrm{acc}}$ is the peak RF voltage amplitude, $\omega_{0}=2\pi f_0$ the angular RF frequency of the cavity, and $\phi_{\mathrm{0}}$ the RF phase. For convenience, the RF voltage can be further decomposed as

\begin{equation} 
V_{\mathrm{RF}}(t) = \underbrace{V_{\mathrm{RF,0}} \cos(\omega_{\mathrm{0}} t)\sin\phi_{\mathrm{0}}}_{V_{\mathrm{RF,acc}}(t)} + \underbrace{V_{\mathrm{RF,0}} \sin(\omega_{\mathrm{0}} t)\cos\phi_{\mathrm{0}}}_{V_{\mathrm{RF,foc}}(t)}, 
\label{eq:V_RF_split} 
\end{equation}
\noindent
where $V_{\mathrm{RF,acc}}(t)$ and $V_{\mathrm{RF,foc}}(t)$ are the accelerating and focusing components of the RF voltage, respectively.

The longitudinal beam-induced voltage is calculated by convolving the longitudinal wake potential with the charge distribution of the bunch as~\cite{Wilson1989}

\begin{equation}
V_{b}(t) = \int_{-\infty}^{t} W_z(t - t') \lambda(t') \mathrm{d}t'.
\label{eq:Vb_def}
\end{equation}

\noindent
This expression is fully general and applies to both pencil beam and finite-transverse-size wake potentials. 

Considering the wake decomposition introduced in Eq.~\eqref{eq:Wtot}, the resonant component of the beam-induced voltage can be written as

\begin{equation}
V_{b\mathrm{,eig}}(t)\!=\!V_{b\mathrm{,FM}}(t)\!+\!\sum_n V_{b\mathrm{,HOM},n}(t)
\label{eq:V_b_tot_dec}
\end{equation}
\noindent
\noindent
where $V_{b\mathrm{,FM}}(t)$ is the voltage induced by the FM and $V_{b\mathrm{,HOM},n}$ is the induced voltage contribution for the $n$-th cavity HOM. Only monopole HOMs with frequencies below the maximum excitable frequency of the beam, determined by the bunch length, are included in the summation. Each term in Eq.~\eqref{eq:V_b_tot_dec} is obtained by applying the general convolution defined in Eq.~\eqref{eq:Vb_def} to the corresponding resonant components of the longitudinal wake potential introduced in Eq.~\eqref{eq:Wtot}.

For the pre-merging section, the total voltage induced by the 21-bunch train is approximated by summing the single-bunch wake voltage contributions of the individual bunches. This is valid since the bunch spacing of 2.84~ns~\cite{taylor2025mucol} is small compared to the FM and HOM wake decay, which are calculated to be on the order of tens of \textmu s. This provides a conservative estimate, as it neglects any partial voltage cancellation among the HOMs. In the post-merging section, the total induced voltage is computed by applying Eq.~\eqref{eq:Vb_def} to the single merged bunch distribution. 

\begin{figure}[hbt!]
    \includegraphics[width=0.48\textwidth]{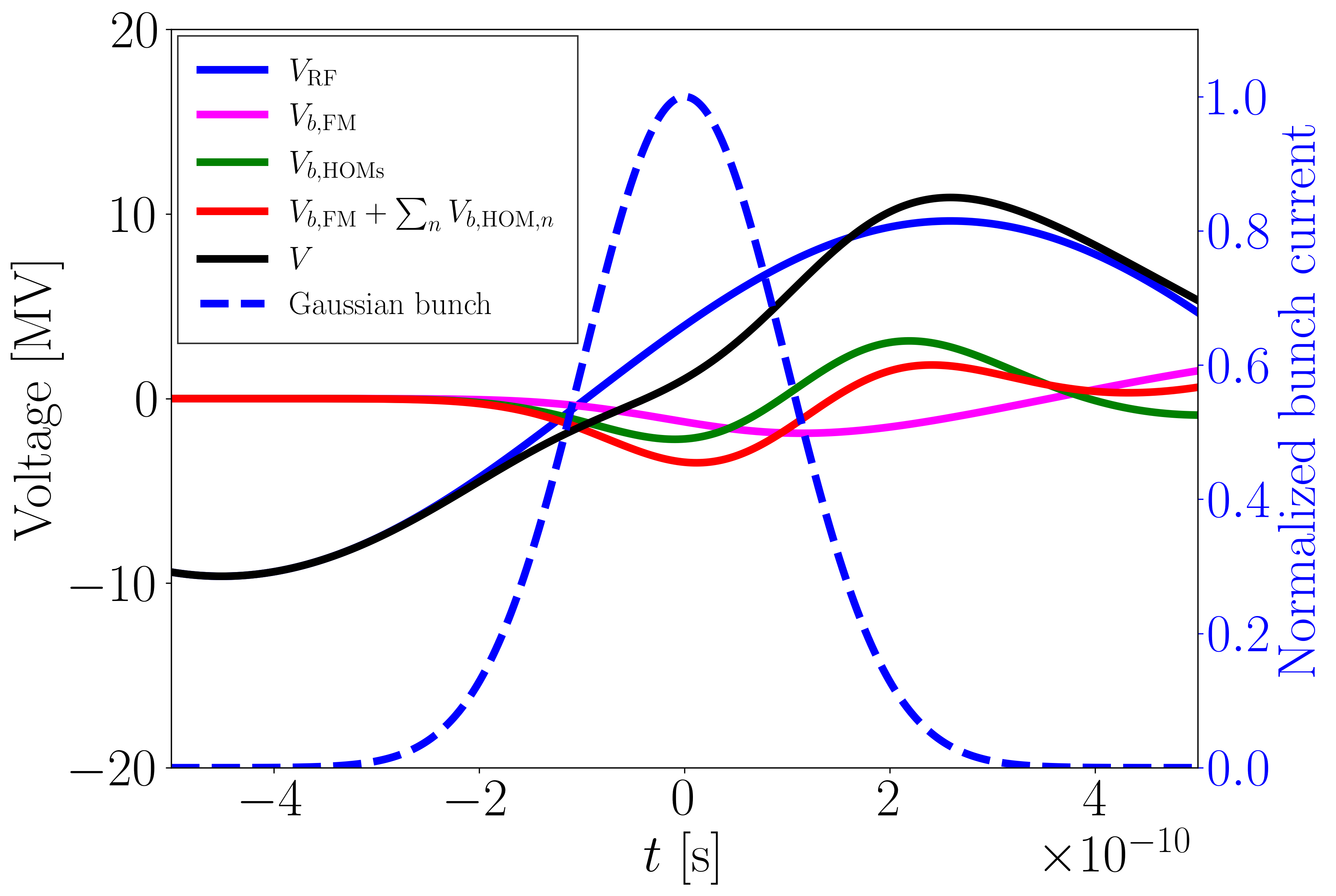}
    \caption{Resonant beam-induced voltage decomposition and input RF voltage for the stage-B5 multi-cell cavity with Be windows. The normalized bunch current distribution is shown on the secondary axis.}
    \label{fig:Voltage_dec_B5_multi_cell}
\end{figure}

Figure~\ref{fig:Voltage_dec_B5_multi_cell} illustrates the computed resonant components of the beam-induced voltage for the stage-B5 multi-cell cavity with Be-windows and the input RF voltage, together with the normalized Gaussian bunch charge distribution used in the simulations. 

\noindent

The analysis of this cavity shows that at the synchronous particle position ($t=0$), corresponding to the peak of the charge distribution, the magnitude of the trapped HOM voltage is approximately double that of the FM ($V_{b,\mathrm{HOMs}} \approx -2.2$~MV compared to $V_{b,\mathrm{FM}}~\approx~-1.2$~MV). This decomposition confirms that the HOMs are the primary source of beam-loading voltage in the short-range wakefield regime, a direct consequence of the Be windows trapping parasitic energy within the cell structure. Specifically, this single-bunch wake is dominated by the lowest-order longitudinal monopole HOMs, namely the $\mathrm{TM}_{020}$ and $\mathrm{TM}_{021}$ modes, due to their higher $(R/Q)_n$ values. Furthermore, because the bunch transit time $t$ is much shorter than the modal decay time $\tau_n$ ($t \ll \tau_n$, yielding $e^{-t/\tau_n} \approx 1$), the amplitude of this short-range wake depends strictly on the term $\omega_n (R/Q)_n$ and is independent of $Q_\mathrm{L}$. Consequently, external HOM dampers, which are used to suppress long-range multi-bunch wakes by lowering $Q_\mathrm{L}$, cannot reduce beam-induced voltage in the short-range wake regime for these cavities. Thus, suppressing beam-induced voltage relies on geometric tuning of $(R/Q)_n$ or shifting the resonant frequency. Overall, the results indicate that the total voltage amplitude reduction is driven by cavity modes, necessitating an increase in RF voltage amplitude. These conclusions remain consistent for the subsequent cavities of the rectilinear muon cooling channel under study.

Equations~\eqref{eq:V_seen_beam} and~\eqref{eq:V_RF_split} show that the beam-induced voltage modifies both the amplitude and the local slope of the RF voltage profile experienced by the bunch. If uncompensated, beam loading can lead to longitudinal emittance growth associated with an increased energy spread~\cite{wangler2008rf}. To mitigate these effects, a local beam-loading compensation scheme is implemented for the RF cavities of the rectilinear muon cooling channel.

In this approach, the compensation conditions require that both the total voltage and its time derivative at the bunch center ($t=0$) match the desired loaded RF voltage $V_{\mathrm{RF,L}}(t)$. Specifically, using the decomposition of the RF voltage introduced in Eq.~\eqref{eq:V_RF_split}, the accelerating component of the RF voltage $V_{\mathrm{RF,acc}}(t)$ is adjusted to compensate for the beam-induced voltage at the bunch center, while the focusing component $V_{\mathrm{RF,foc}}(t)$ is modified to restore the required voltage slope at the same time. This leads to the following system of equations:

\begin{equation}
\left\{
\begin{aligned}
&V_{\mathrm{RF}}(t) + V_{{b}}(t) = V_{\mathrm{RF,L}}(t) \\
&\displaystyle \frac{\partial V_{\mathrm{RF}}(t)}{\partial t} + \frac{\partial V_{{b}}(t)}{\partial t} = \frac{\partial V_{\mathrm{RF,L}}(t)}{\partial t}
\end{aligned}
\right. \quad \text{at } t=0
\label{eq:beam_loading_compensation_system}
\end{equation}

\noindent
By satisfying both conditions, the perturbations to the accelerating and focusing components introduced by the beam-induced voltage near $t=0$ are suppressed. As a result, the amplitude of the required RF voltage for muon acceleration increases, while the voltage profile remains linear near the bunch center, thereby mitigating energy-spread and longitudinal phase-space distortion in this region. 

This beam-loading compensation scheme is constrained by its linear nature, requiring the voltage contributions to be independent of the transverse beam offset, as discussed in Sec.~\ref{wakefield_section}. For cavities without Be windows, both the FM and HOM wake potentials are independent of the transverse beam offset. Furthermore, the indirect space-charge contribution is also offset-independent. In this case, the beam voltage induced by these three contributions can be treated as a purely longitudinal effect, and the compensation scheme can be applied to the contribution \(V_{b,\mathrm{FM}} + V_{b,\mathrm{HOM}} + V_{b,\mathrm{ISC}}\). In contrast, for cavities equipped with Be windows, while the FM and HOM contributions remain suitable for linear compensation, the indirect and direct space-charge terms show a nonlinear dependence on the transverse offset of the beam. As a consequence, only the eigenmode-driven contributions, i.e., \(V_{b,\mathrm{FM}} + V_{b,\mathrm{HOM}}\), allow a linear beam-loading compensation in cavities with Be windows. Therefore, beam-loading compensation is applied exclusively to the FM and HOM contributions in cavities with Be windows.

\begin{table*}
	\caption{Main beam loading results and RF figures of merit impacted by beam loading compensation for the multi-cell RF cavities with Be windows in the rectilinear cooling channel. Only the eigenmode-driven contributions, i.e., \(V_{b,\mathrm{FM}} + V_{b,\mathrm{HOM}}\), are compensated. The calculation of the beam-induced voltage accounts for a beam of finite transverse size of $\sigma_{\mathrm{t}}=R_i/3$. Dividing the voltage and power values by $N_{\mathrm{cell}}$ yields the equivalent values for a single-cell RF cavity.}
    \setlength{\tabcolsep}{3.4pt}
		\begin{tabular}{cccccccccccccc}
          \toprule
			& $V_{\mathrm{RF,}0}$ & $|V_b(t=0)|$ & $V_{b,\mathrm{max}}$ &  $|V_b(t=0)|/V_{\mathrm{RF,}0}$ & $V_{b,\mathrm{max}}/V_{\mathrm{RF,}0}$ & $V_{\mathrm{RF,L,0}}$ & $V_{\mathrm{RF,L,0}}/V_{\mathrm{RF,}0}$ & $P_{\mathrm{diss}}^{'}$ & $P_{\mathrm{g}}^{'}$ & $E_{\mathrm{peak,Cu}}^{'}$ & $E_{\mathrm{peak,Be}}^{'}$ \\
			Stage & [MV] & [MV] & [MV] & [1] & [1] & [MV] & [1] & [MW] & [MW] &  [MV/m] & [MV/m]  \\
			\hline
			A1 & 28.31 & 2.72 & 5.35 & 0.10 & 0.19 & 32.44 & 1.15 & 33.44 & 40.12 & 13.43 & 31.37 \\
			A2  & 18.07 & 2.48 & 4.54 & 0.14 & 0.25 & 22.69 & 1.26 & 27.38 & 32.85 & 29.19 & 33.28 \\
            A3  & 13.47 & 4.61 & 8.91 & 0.34 & 0.66 & 13.75 & 1.02 & 10.70 & 12.84 & 21.22 & 32.14 \\
            A4  & 10.87 & 5.44 & 9.02 & 0.50 & 0.83 & 22.32 & 2.05 & 37.32 & 44.78 & 57.23 & 65.36 \\
            B1  & 26.32 & 2.22 & 3.92 & 0.08 & 0.15 & 30.59 & 1.16 & 21.71 & 26.05 & 14.14 & 24.70 \\
            B2  & 20.62 & 1.49 & 2.64 & 0.07 & 0.13 & 23.38 & 1.13 & 18.03 & 21.63 & 17.29 & 24.66 \\
            B3  & 17.01 & 1.53 & 2.31 & 0.09 & 0.14 & 20.45 & 1.20 & 23.52 & 28.22 & 31.47 & 29.37 \\
            B4  & 13.97 & 1.18 & 1.72 & 0.08 & 0.12 & 16.80 & 1.20 & 17.00 & 20.40 & 33.35& 27.45 \\
            B5  & 9.63 & 1.96 & 3.42 & 0.20 & 0.35 & 13.28 & 1.38 & 11.26 & 13.51 & 33.28 & 30.40 \\
            B6  & 9.60 & 1.60 & 2.80 & 0.17 & 0.29 & 12.88 & 1.34 & 13.56 & 16.27 & 44.64 & 35.56 \\
            B7  & 9.79 & 1.48 & 2.56 & 0.15 & 0.26 & 13.11 & 1.34 & 14.12 & 16.94 & 47.20 & 34.79 \\
            B8  & 9.34 & 1.35 & 2.36 & 0.14 & 0.25 & 12.58 & 1.35 & 12.95 & 15.54 & 46.69 & 30.82 \\
            B9  & 10.26 & 1.25 & 2.12 & 0.12 & 0.21 & 13.47 & 1.31 & 14.79 & 15.75 & 50.60 & 30.56 \\
            B10  & 8.63 & 1.07 & 1.83 & 0.12 & 0.21 & 11.39 & 1.32 & 10.51 & 12.62 & 42.94  & 24.23 \\
		\toprule
        \end{tabular}
          
	\label{tab:beam_loading_cavities_windows}
\end{table*}

\begin{table*}
	\caption{Main beam loading results and RF figures of merit impacted by beam loading compensation for the stage-B5 multi-cell cavity without windows. The combination of eigenmode-driven and indirect space-charge contributions, i.e., \(V_{b,\mathrm{FM}} + V_{b,\mathrm{HOM}}+ V_{b,\mathrm{ISC}}\), is compensated. Dividing the voltage and power values by $N_{\mathrm{cell}}$ yields the equivalent values for a single-cell RF cavity.}
    \setlength{\tabcolsep}{3.4pt}
		\begin{tabular}{ccccccccccccc}
          \toprule
			& $V_{\mathrm{RF,}0}$ & $|V_b(t=0)|$ & $V_{b,\mathrm{max}}$ &  $|V_b(t=0)|/V_{\mathrm{RF,}0}$ & $V_{b,\mathrm{max}}/V_{\mathrm{RF,}0}$ & $V_{\mathrm{RF,L,0}}$ & $V_{\mathrm{RF,L,0}}/V_{\mathrm{RF,}0}$ & $P_{\mathrm{diss}}^{'}$ & $P_{\mathrm{g}}^{'}$ & $E_{\mathrm{peak,Cu}}^{'}$ \\
			Stage & [MV] & [MV] & [MV] & [1] & [1] & [MV] & [1] & [MW] & [MW] &  [MV/m] \\
			\hline
			
            B5  & 9.63 & 0.87 & 1.62 & 0.09 & 0.17 & 10.29 & 1.07 & 14.58 & 17.50 & 52.20  \\
            
		\toprule
        \end{tabular}
          
	\label{tab:beam_loading_cavities_without_windows}
\end{table*}

\begin{figure}[hbt!]
    \centering

    \begin{subfigure}{0.453\textwidth}
        \centering
        \includegraphics[width=\textwidth]{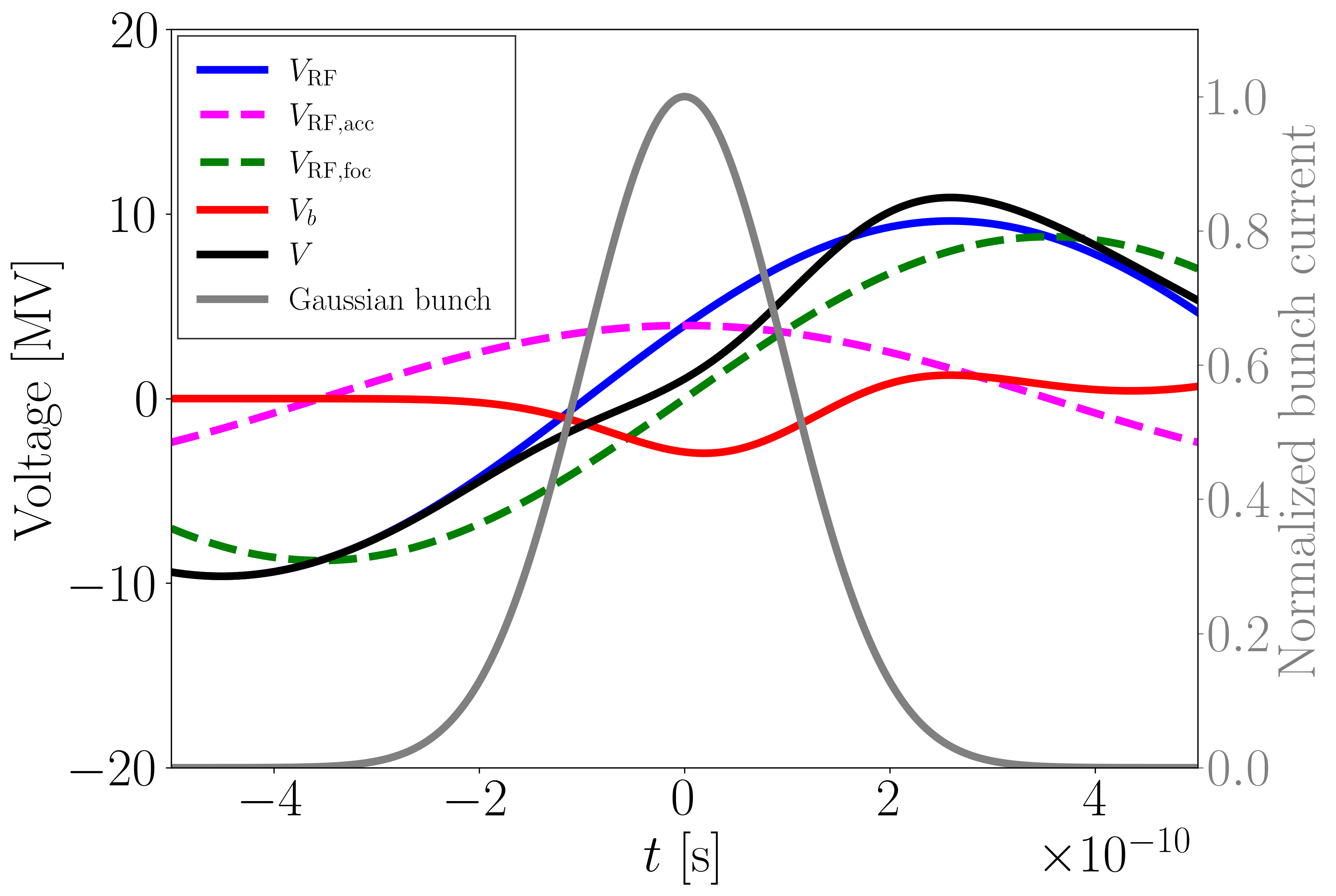}
        \caption{}
        \label{fig:Beam_loading_FM_HOMs_before_compensation}
    \end{subfigure}

    \vspace{0.5cm}

    \begin{subfigure}{0.453\textwidth}
        \centering
        \includegraphics[width=\textwidth]{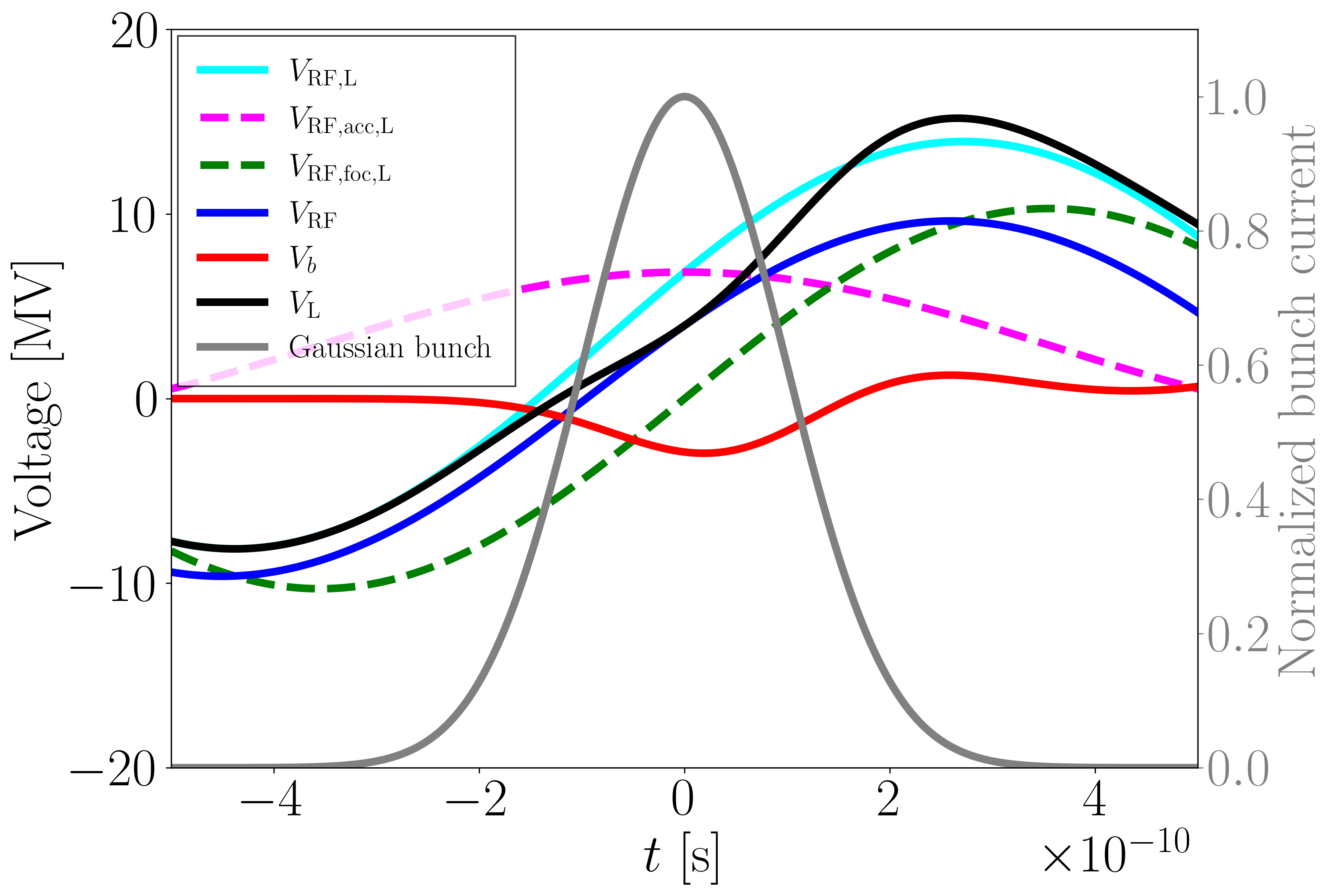}
        \caption{}
        \label{fig:Beam_loading_FM_HOMs_after_compensation}
    \end{subfigure}

    \caption{Beam loading and local compensation in the stage-B5 multi-cell cavity. The total voltage experienced by the beam and the RF voltage components are shown before (a) and after (b) beam loading compensation, alongside the beam-induced voltage from the FM and HOMs. The normalized bunch current distribution is illustrated on the secondary axis.}
    \label{fig:Beam_loading_FM_HOMs}
\end{figure}

The effect of beam loading and its local compensation in the stage-B5 multi-cell cavity is shown in Fig.~\ref{fig:Beam_loading_FM_HOMs}. The total voltage seen by the beam, together with the RF voltage components, is shown before and after compensation in Fig.~\ref{fig:Beam_loading_FM_HOMs_before_compensation} and in Fig.~\ref{fig:Beam_loading_FM_HOMs_after_compensation}, respectively. The beam-induced voltage arising from the combination of FM and HOMs is also illustrated with the normalized bunch current distribution.  

Comparing these plots confirms that the used compensation scheme effectively restores the total voltage profile. As shown in Fig.~\ref{fig:Beam_loading_FM_HOMs}(b), near the bunch center ($t=0$), the net voltage experienced by the beam $V_{\mathrm{L}}$ (black curve) successfully matches the amplitude and local slope of the design RF voltage $V_{\mathrm{RF}}$ (blue curve). Under heavy beam loading, the RF generator must provide an increased cavity voltage $V_{\mathrm{RF,L}}$ to compensate for the beam-induced voltage and preserve the operating accelerating gradient. This voltage increase results in a quadratic increase in the power dissipated on the cavity walls, defined by $P_{\mathrm{diss}}' = P_{\mathrm{diss}} \cdot (V_{\mathrm{RF}}/V_{\mathrm{RF,L}})^2$, and a linear rise in the peak surface electric field, $E_{\mathrm{peak}}'=E_{\mathrm{peak}}\cdot(V_{\mathrm{RF}}/V_{\mathrm{RF,L}})$, where $P_{\mathrm{diss}}'$ and $E_{\mathrm{peak}}'$ are the dissipated power and the surface electric field after compensation, respectively. These effects impose higher requirements on the available RF power and the electrical breakdown limits of the cavities. The main results of the beam loading analysis for the cavities with Be windows of each stage of the rectilinear cooling channel and the main RF figures of merit impacted by the beam loading compensation are reported in Table~\ref{tab:beam_loading_cavities_windows}. The calculation of the beam-induced voltage assumes a beam of finite transverse size of $\sigma_{\mathrm{t}}=R_i/3$~\cite{RogersPriv}, where $R_i$ is the iris radius of the cavity under consideration. The results show that, depending on the cooling stage, the RF voltage required after beam-loading compensation increases by approximately 15\% up to nearly a factor of two in the most demanding case of stage A4. The corresponding peak surface electric fields are much higher than those of the uncompensated ones in Table~\ref{tab:RF_parameters}. 

To further assess the impact of the cavity geometry on beam-loading compensation, a comparison was carried out for the stage-B5 cavity without Be windows. The results are reported in Table~\ref{tab:beam_loading_cavities_without_windows}. The data indicate that, although the absence of windows lowers the required loaded RF voltage for beam loading compensation, it shifts the cavity operation towards higher surface electric fields and higher RF power dissipation. Considering the stage-B5 cavity as a representative example, removing the Be windows reduces $V_{\mathrm{RF,L}}$ from 13.28~MV to 10.29~MV once the respective beam loading compensations are applied. However, the dissipated power rises from 11.26~MW to 14.58~MW, and the peak surface electric field increases from 33.28~MV/m to 52.20~MV/m, thereby underlining better performance of the cavity with Be windows. Similar behavior is expected for the other cooling stages, given the comparable cavity geometries and operating conditions. Consequently, the use of Be windows in the muon cooling cavities appears to be a critical design feature for mitigating the risk of RF breakdown while maintaining the high gradients necessary for efficient muon cooling.

\vspace{1.1em}

\section{Conclusion}\label{sec:5}

This paper presents a comprehensive RF design and beam-induced effects analysis of high-gradient normal-conducting copper cavities for the rectilinear cooling channel of the muon collider. This study establishes a physics-based design framework that integrates RF cavity optimization with beam-loading compensation.

RF elliptical cavities made of copper and closed at the irises with Be windows were developed and optimized in accordance with the beam dynamics specifications of the recent cooling lattice. Each cavity was designed to maintain a high transit time factor per cell and to maximize the $R/Q$ within the geometric constraints imposed by the lattice. The peak surface electric field was also lowered to minimize the risk of RF breakdown. The RF figures of merit and power requirements were calculated for different cavity configurations and cooling stages, demonstrating compatibility with pulsed normal-conducting operation.

A comparison of single-cell cavities with and without Be windows demonstrated the essential role of the windows in achieving acceptable RF characteristics for muon cooling. The cell configuration without windows yields a significant reduction in $R/Q$, increased dissipated power, and higher peak surface fields, ultimately making such a concept unsuitable. Therefore, within the range of configurations studied, the inclusion of Be windows appears to be a necessary design choice to satisfy acceleration efficiency, power, and breakdown constraints simultaneously.

Wakefields and beam loading effects in the developed RF cavities were systematically analyzed using a hybrid frequency-time domain numerical method developed to isolate the space-charge contribution from the total wake potential. The transverse dependence of the wake potential was studied for cavities with and without Be windows, revealing different behaviors. In cavities equipped with windows, a radial dependence in the wake contributions associated with the fundamental mode, higher-order modes, and indirect space-charge components is observed. In contrast, for cavities without windows, these contributions are essentially independent of the transverse offset, and the offset sensitivity of the total wake potential is mainly due to the direct space-charge term.

An important implication of these results is that, in cavities with Be windows, only the eigenmode-driven contributions were treated for a linear beam-loading compensation in these cavities. To reconstruct the longitudinal eigenmode wake potential more realistically, finite transverse beam-size effects were included in the proposed eigenmode analytical formulation. The analysis reveals that the transverse beam size reduces the wake potential. The analytical model was benchmarked against full three-dimensional PIC simulations, demonstrating very good agreement between both approaches.

A local beam-loading compensation scheme was then presented and implemented. In this approach, the accelerating component of the RF voltage is adjusted to cancel the beam-induced voltage at the bunch center, while the focusing component is modified to restore the required voltage slope. For the cavity design with windows, beam-loading compensation of the FM and HOM was verified numerically. Depending on the cooling stage, the required RF voltage after compensation increases from approximately 15\% up to a factor of two in the most demanding case, resulting in a significant increase in the surface fields and required RF power. These effects can be directly mitigated by reducing beam loading through geometric optimization of $(R/Q)_n$ or frequency shifting.
While this local scheme successfully mitigates single-bunch effects, compensating for a multi-bunch train is beyond the scope of this work and would require extending the algorithm to handle potential long-range wakefield buildup. In the future, full beam dynamics studies will be required to evaluate the effect of beam loading and its compensation on the performance of the muon cooling channel.

Overall, this work provides quantitative design guidelines for RF systems in the rectilinear cooling channel of a future muon collider. Cavity designs were optimized, key RF parameters were quantified across configurations and stages, beam-induced effects were analyzed, and a viable local beam-loading compensation strategy was developed. Ultimately, this work establishes a practical and technically consistent RF cavity design framework for high-intensity muon cooling applications.

\section{ACKNOWLEDGEMENTS}
The authors would like to thank Prof. Ursula van Rienen and Dr. Heiko Damerau for their valuable feedback on the manuscript. The research presented here has been performed within the framework of the International Muon Collider Collaboration (IMCC). This work is funded by the European Union (EU). Views and opinions expressed are, however, those of the authors only and do not necessarily reflect those of the EU or European Research Executive Agency (REA). Neither the EU nor the REA can be held responsible for them. The authors also acknowledge support by the US Department of Energy, Office of Science under Awards DE‐SC0014664 and DE-SC0021928, and Contract DE-AC02-05CH11231.


\end{document}